\documentclass[12pt]{article}
\usepackage{graphicx,  amsmath,amssymb,latexsym,psfrag,epsfig,subfigure,euscript, multirow,color}
\allowdisplaybreaks

\def\rd{\mathrm{d}}

\def\lepone{{\sc lep} 1 }
\def\leptwo{{\sc lep} 2 }
\def\first{{1$^{\mathrm{st}}$}}
\def\second{{2$^{\mathrm{nd}}$}}
\def\third{{3$^{\mathrm{rd}}$}}
\def\fourth{{4$^{\mathrm{th}}$}}
\def\wt{\widetilde}

\newcommand\tf{\widetilde{s}_f}

\definecolor{darkred}{rgb}{0.6,0.0,0.0}
\definecolor{darkblue}{rgb}{0.0,0.0,0.8}
\definecolor{darkgreen}{rgb}{0.0,0.5,0.0}
\definecolor{brown}{rgb}{0.0,0.0,0.0}
\newcommand{\red}{\color{darkred}}
\newcommand{\blue}{\color{darkblue}}

\newcommand{\rLO}{ {\red L_1} }
\newcommand{\rLT}{ {\red L_2} }
\newcommand{\rLM}{ {{\red L}} }
\newcommand{\rL}{ {{\red L}} }

\newcommand{\cone}{ {\blue c^S_1} }
\newcommand{\ctwo}{ {\blue c^S_2} }
\newcommand{\ctwoL}{ {\blue c^S_{2L}} }
\newcommand{\ctwoQ}{ {\blue c^S_{2Q}} }
\newcommand{\ctwor}{ {\blue c^S_{2\rho}} }
\newcommand{\ctwoz}{ {\blue c^S_{2\zeta}} }

\newcommand{\ctwocf}{ {\blue c^S_{2C_F}} }
\newcommand{\ctwoca}{ {\blue c^S_{2C_A}} }
\newcommand{\ctwonf}{ {\blue c^S_{2n_f}} }
\newcommand{\ctworcf}{ {\blue c^S_{2\rho C_F}} }
\newcommand{\ctworca}{ {\blue c^S_{2\rho C_A}} }
\newcommand{\ctwornf}{ {\blue c^S_{2\rho n_f}} }
\newcommand{\ctwozcf}{ {\blue c^S_{2\zeta C_F}} }
\newcommand{\ctwozca}{ {\blue c^S_{2\zeta C_A}} }
\newcommand{\ctwoznf}{ {\blue c^S_{2\zeta n_f}} }
\newcommand{\rmin} { {\rho_{\text{min}}}}

\def\cdt{ \hspace{-0.1em} \cdot \hspace{-0.1em} }

\newcommand{\Lnp}{ {\Lambda_{\mathrm{NP}}}}
\newcommand{\LNP}{ {\Lambda_{\mathrm{NP}}}}
\newcommand{\LQCD}{ {\Lambda_{\mathrm{QCD}}}}

\def\nn{\nonumber}

\begin{document}

\begin{titlepage}

\vspace{0.2cm}
\begin{center}
\Large\bf
Resummation of heavy jet mass and \\
comparison to LEP data
\end{center}

\vspace{0.2cm}
\begin{center}
{\sc Yang-ting Chien and Matthew D. Schwartz}\\
\vspace{0.4cm}
Center for the Fundamental Laws of Nature\\
Harvard University\\
Cambridge, MA 02138, U.S.A.
\end{center}

\vspace{0.2cm}
\begin{abstract}
\vspace{0.2cm}
\noindent 

The heavy jet mass distribution in $e^+e^-$ collisions is computed to 
next-to-next-to-next-to leading logarithmic (N${}^3$LL) and
next-to-next-to leading fixed order accuracy (NNLO). The singular terms predicted from the
resummed distribution are confirmed by the fixed order distributions
allowing a precise extraction of the unknown soft function coefficients.
A number of quantitative
and qualitative comparisons of heavy jet mass and the related thrust distribution are made. 
From fitting to ALEPH data, a value of $\alpha_s$ is extracted, $\alpha_s(m_Z)=0.1220 \pm 0.0031$,
which is larger than, but not in conflict with, the corresponding value for thrust. A weighted average of the two produces
$\alpha_s(m_Z) = 0.1193 \pm 0.0027$, consistent with the world average. 
A study of the non-perturbative corrections shows that
the flat direction observed for thrust
between $\alpha_s$ and a simple non-perturbative shape parameter is not lifted in combining with heavy jet mass.
The Monte Carlo treatment of hadronization gives qualitatively different results for thrust and heavy jet mass,
and we conclude that it cannot be trusted to add power corrections to the event shape distributions at this accuracy.
Whether a more sophisticated effective field theory approach to power corrections can reconcile the
thrust and heavy jet mass distributions remains an open question.
\end{abstract}
\vfil

\end{titlepage}

\section{Introduction}
Event shapes in $e^+e^-$ collisions provide some of the best ways to test QCD and the standard model. 
At high energies, where QCD is perturbative, event shapes lead to some of the world's
most precise measurements of the strong coupling constant $\alpha_s$.
Recently, a number of theoretical advances have led to renewed interest in event shapes and the $\alpha_s$ measurements. First,
the NNLO fixed order Feynman diagrams were calculated~\cite{GehrmannDeRidder:2005cm,GehrmannDeRidder:2007bj, GehrmannDeRidder:2007hr,Weinzierl:2008iv}.
This allowed the prediction of all event shapes
to order $\alpha_s^3$. Second, advances~\cite{Fleming:2007qr,Schwartz:2007ib,Becher:2006qw}
 in Soft-Collinear Effective Theory (SCET)~\cite{Bauer:2000yr,Bauer:2001yt,Beneke:2002ph} 
have allowed resummation of the large logarithmic corrections to thrust to N${}^3$LL accuracy~\cite{Becher:2008cf}. 
Previous calculations were at NLO~\cite{Ellis:1980wv} and NLL~\cite{Catani:1992ua}.
Very recently, a full effective field theory analysis of a single event shape, thrust, has been completed~\cite{Abbate:2010vw},
including additionally non-perturbative considerations. The resulting $\alpha_s$ extraction is competitive
with the PDG world average~\cite{Yao:2006px}, however it differs significantly from measurements using lattice 
QCD and $\tau$ decays (see~\cite{Bethke:2009jm} for a review).

Although the $\alpha_s$ measurement with thrust is extremely precise, there are many reasons to study additional event
shapes as well. The main advantage is that there may be systematic effects in a single event shape pulling $\alpha_s$
in a certain direction, which are not universal. 
In fact, as observed in~\cite{Dissertori:2009ik}
from an NLL+NNLO analysis, there seem to be two classes of event shapes, the first including thrust, the $C$-parameter
and total jet broadening, while the second includes heavy jet mass, wide jet broadening and the two-to-three jet transition parameter $y_{3}$.
The values of $\alpha_s$ extracted from the two classes at NLL+NNLO tend
to have around a 5\% systematic difference which the authors attribute to missing higher order corrections.
In a recent world average of $\alpha_s$~\cite{Bethke:2009jm}, the thrust measurement using SCET was not included
because of a concern over precisely this kind of systematic uncertainty. In this paper, we correct that concern with
a N${}^3$LL calculation of an event shape from the second class, heavy jet mass.

In addition to being useful for measuring $\alpha_s$, the heavy jet mass distribution allows us to explore
other aspects of resummation. Indeed, there are few hadronic observables which have been calculated this
accurately and for which there is data. Heavy jet mass involves a soft function which cannot be written
in terms of only a single scale. These types of soft functions promise to play an important
role in resummation at hadron colliders and only beginning to be explored~\cite{Becher:2009th,Ellis:2010rw,Bauer:2010vu}.
 We will discuss constraints on the 
soft function, and perform a numerical study of the parts that are not known, similar to what
was done in~\cite{Becher:2008cf} and \cite{Hoang:2008fs}.

Hadronization is another issue which having a second event shape may help understand. In the fit
to $\alpha_s$ with the thrust distribution~\cite{Becher:2008cf}, it was observed that a decrease in $\alpha_s$ could 
be compensated for with a single non-perturbative parameter with only a small effect on the $\chi^2$ of
the fit. Having another observable for which the same non-perturbative parameters can be fit
can possibly remove this flat direction. The hadronization issue is also important for Monte Carlo simulations.
With a more accurate theoretical calculations, we can explore whether the approximations in {\sc pythia}~\cite{Sjostrand:2006za} allow 
for an adequate description of thrust and heavy jet mass simultaneously.

As a brief outline of our findings, we begin in Section~\ref{sec:dist} with an overview of the SCET distributions. The hemisphere soft function
is studied and singular terms in the  heavy jet mass distribution are compared to the fixed order calculation
 in Section~\ref{sec:soft}. We found a mild inconsistency
with the analytic results from SCET and the numerical calculations of the NLO and NNLO distributions. 
After completing the original study, we were able to resolve this  inconsistency,
which was due to Monte Carlo convergence problems,
by taking a very low  numerical infrared cutoff, as discussed in a
note added at the end of this paper.
In Section~\ref{sec:fit} we fit for $\alpha_s$. The fit for heavy jet mass to the {\sc lep} data from {\sc aleph}~\cite{Heister:2003aj}
 leads to 
$\alpha_s(m_Z) = 0.1220 \pm 0.0031$. This value is higher than the value from thrust using exactly the
same technique, $\alpha_s(m_Z) = 0.1175 \pm 0.0026$. Assuming 100\% correlation gives an average value of $\alpha_s(m_Z) = 0.1193 \pm 0.0027$ which is very 
close to the recent average in~\cite{Bethke:2009jm}. We also find that convergence of the perturbation series for heavy jet mass with resummation is, like for thrust,
significantly better than the convergence of the fixed order calculation. 
In Section~\ref{sec:NP}, a comparison of the data to {\sc pythia} shows that while {\sc pythia} agrees
with the thrust data almost perfectly, it has trouble matching the heavy jet mass distribution. Moreover, the hadronization corrections in {\sc pythia} move
the curve in the wrong direction for heavy jet mass. Concluding that the Monte Carlo hadronization model is incompatible with the high precision theoretical calculation,
we explore non-perturbative corrections in SCET with a simple shape function. We find that to the order we are working, this simple shape function cannot
simultaneously describe the thrust and heavy jet mass distributions. We provide an expanded summary, discussion and comparison
to previous results in Section~\ref{sec:conc}.

\section{Thrust and Heavy Jet Mass in SCET \label{sec:dist}}
Thrust and heavy jet mass are defined as follows. One first finds the thrust axis, through
\begin{equation}
  T = \max_{\mathbf{n}} \frac{\sum_i | \mathbf{p}_i \cdot \mathbf{n}
  |}{\sum_i | \mathbf{\mathbf{p}}_i |}\,,
\end{equation}
where the sum is over all momentum 3-vectors $\mathbf{p}_i$ in the event,
and the maximum is over all unit 3-vectors ${\mathbf{n}}$. We use
$\tau = 1-T$ to measure thrust. Once the thrust axis is known, it can be used
to split the event into two hemispheres. We define $P_L^\mu$ and $P_R^\mu$ to be the four
momenta of the sum of all the radiation going into each hemisphere, and $M_L = \sqrt{P_L^2}$
and $M_R = \sqrt{P_R^2}$ to be the hemisphere masses. 
Heavy jet mass $\rho$ is defined as
the larger of the two hemisphere masses squared, normalized to the the center of mass energy $Q$,
\begin{equation}
  \rho \equiv \frac{1}{Q^2} \max(M_L^2 , M_R^2) \, .
\end{equation}
When  $\rho$ is small, $\tau$ is also small, both hemisphere masses are small,
and the event appears to have two back-to-back pencil-like jets. In this threshold limit, the thrust
axis aligns with the jet axis and $Q^2\tau$ approaches the sum of the two hemisphere masses squared
$M_L^2 + M_R^2  = Q^2\tau + {\mathcal O}(\tau^2)$.

It follows that both thrust, up to power corrections, and heavy jet mass can be written
as integrals over the doubly differential hemisphere mass distribution:
\begin{align}
  \frac{\rd \sigma}{\rd \tau} &=Q^2 \int \frac{\rd^2 \sigma}{\rd M_L^2\rd M_R^2} \delta(Q^2\tau - M_L^2 - M_R^2)\, ,\\
  \frac{\rd \sigma}{\rd \rho} &=Q^2 \int \frac{\rd^2 \sigma}{\rd M_L^2 \rd M_R^2}
\left[ \delta(Q^2\rho - M_L^2)\theta(M_L^2 - M_R^2) +  \delta(Q^2\rho - M_R^2)\theta(M_R^2 - M_L^2) \right]\label{rhofirst} \,.
\end{align}

In SCET, the doubly differential hemisphere mass distribution is calculable in the threshold limit.
The appropriate factorization theorem in SCET was first derived in~\cite{Fleming:2007qr} for the
related process of $t\bar{t}$ production. This theorem was then shown to allow for the calculation of
event shapes in~\cite{Schwartz:2007ib}, where matched and resummed thrust and heavy jet mass distributions in SCET were first
presented. Previously, resummation of heavy jet mass was only possible at NLL accuracy~\cite{Catani:1992ua}.
The first event shape resummed to N${}^3$LL was thrust, in~\cite{Becher:2008cf}. Monte Carlo based hadronization corrections
were included in~\cite{Kaplan:2008pt} to produce a strong model-independent gluino mass bound. Recently,
power corrections for thrust were studied within the effective field theory approach in~\cite{Abbate:2010vw}.

The factorization theorem allows us to write the hemisphere mass distribution as
\begin{equation} \label{hemidist}
  \frac{1}{\sigma_0} \frac{\rd^2 \sigma_2}{\rd M_L^2 \rd M_R^2} = H(Q^2, \mu) \int \! 
\rd  k_L \rd k_R \,
J (M^2_L- Q k_L, \mu) \,J (M^2_R - Q k_R, \mu)\, S (k_L, k_R, \mu) \,.
\end{equation}
The subscript on $\sigma_2$ is a reminder that this expression holds in the two-jet region.
Here, $H(Q^2,\mu)$ is the hard function. It is calculated in matching SCET to QCD and contains information
about the modes of QCD that are not in SCET. $J(p^2,\mu)$ is a jet function. It is derived in a matching
calculation from a theory with soft and collinear modes to a theory with just soft modes. 
The hard function was calculated in~\cite{Becher:2006mr} and the jet functions in~\cite{Becher:2006qw}.
Finally, $S(k_L,k_R,\mu)$ is the hemisphere soft function which is derived from integrating out the remaining
soft modes.

The doubly differential hemisphere mass distribution, Eq.~\eqref{hemidist}, is observable, and therefore must be independent of renormalization
group scale $\mu$. Demanding $\mu$-independence leads to a renormalization group equation which is easiest to express in Laplace space,
where the convolutions turn into products. The Laplace
transform is defined by
\begin{equation}  \label{lapdef}
 \widetilde{f}({\red \nu_L}, {\red \nu_R})=
  \int \rd M_L^2 \rd M_R^2 
e^{-{\red \nu_L} M_L^2}
e^{-{\red \nu_R} M_R^2}
f( M_L^2, M_R^2)
\end{equation}
which can be applied to the differential cross section and to the jet and soft functions
separately.  We generally express the Laplace transformed distributions
as functions of $\rLO = \ln(\mu {\red  \nu_L} e^{\gamma_E})$ and $\rLT = \ln(\mu {\red \nu_R} e^{\gamma_E})$.
Since the entire $\mu$-dependence of the hard and jet functions is known, the $\mu$-dependence of the soft function is completely
fixed by renormalization group invariance (see~\cite{Becher:2008cf} for more details). The result is that the  hemisphere soft function itself factorizes into the form~\cite{Fleming:2007qr,Schwartz:2007ib,Fleming:2007xt,Hoang:2008fs}
\begin{equation}
  \widetilde{s}({\rLO},{\rLT},\mu) =
\widetilde{s}_\mu({\rLO},\mu)
\widetilde{s}_\mu({\rLT},\mu) 
 \tf( {\rLO} - {\rLT}) \,,
\end{equation}
where all the $\mu$-dependence is contained in the function $\wt{s}_\mu( \rL,\mu )$
 which 
is known to N${}^3$LL accuracy.
Since ${\rLO} - {\rLT} = \ln({\red \nu_L}/{\red \nu_R})$, the function $\tf( {\rLO} - {\rLT})$ 
is $\mu$-independent. 
We discuss the soft function more in the next section.

Putting together the hard and jet functions with the soft function written in this way
produces an analytic expression for the doubly differential jet mass distribution.
For thrust, the result is~\cite{Becher:2008cf}
\begin{multline} \label{scetdist}
\frac{1}{\sigma_0} R^{\tau}_2(\tau) 
=
\frac{1}{\sigma_0} \int_0^{\tau}\rd \tau' \frac{\rd\sigma_2}{\rd\tau'}\\
=
\exp\left[ 4S(\mu_h,\mu_j)+4S(\mu_s,\mu_j)-2A_H(\mu_h,\mu_s)+4A_J(\mu_j,\mu_s)\right] \left(\frac{Q^2}{\mu_h^2}\right)^{-2A_\Gamma(\mu_h,\mu_j)} \\
\times H(Q^2,\mu_h)\, 
\left[\widetilde j\Big( \ln\frac{\mu_s Q}{\mu_j^2}+\partial_\eta,\mu_j\Big)\right]^2\, 
\widetilde s_T(\partial_{\eta},\mu_s)
\left[\left(\frac{\tau Q}{\mu_s}  \right)^{\eta} \frac{e^{-\gamma_E \eta}}{\Gamma(\eta+1)}\right]
\,,
\end{multline}
with $\eta = 4 A_{\Gamma} (\mu_j, \mu_s)$ and the thrust soft function $\widetilde{s}_T(\rL,\mu)$ is defined by
\begin{equation}
\widetilde s_T(\rL,\mu) = 
  \widetilde{s}(\rL,\rL,\mu) =[\widetilde{s}_\mu(\rL,\mu)]^2  \tf(0) \,.
\end{equation}
The definitions of the RG kernels $A_\Gamma(\nu,\mu)$ and $S(\nu,\mu)$ as well as the fixed order hard and jet functions, $H(Q^2,\mu)$
and $\wt{j}(\rL,\mu)$ and their anomalous dimensions can be found in~\cite{Becher:2008cf}.
Note that only one value of the unknown function $\tf(\rL)$ is required for thrust. 

For heavy jet mass, the distribution is similar
\begin{multline}  \label{hjmdist}
\frac{1}{\sigma_0} R^{\rho}_2(\rho) 
=
\frac{1}{\sigma_0} \int_0^{\rho}\rd \rho' \frac{\rd\sigma_2}{\rd\rho'}\\
=
\exp\left[ 4S(\mu_h,\mu_j)+4S(\mu_s,\mu_j)-2A_H(\mu_h,\mu_s)+4A_J(\mu_j,\mu_s)\right] \left(\frac{Q^2}{\mu_h^2}\right)^{-2A_\Gamma(\mu_h,\mu_j)} 
\, \\
\times H(Q^2,\mu_h)
\widetilde j\Big( \ln\frac{\mu_s Q}{\mu_j^2}+\partial_{\eta_1},\mu_j\Big)\,
\widetilde j\Big( \ln\frac{\mu_s Q}{\mu_j^2}+\partial_{\eta_2},\mu_j \Big)\,
\widetilde s_\mu(\partial_{\eta_1},\mu_s)
\widetilde s_\mu(\partial_{\eta_2},\mu_s)
\left(\frac{\rho Q}{\mu_s}  \right)^{{\eta_1}+{\eta_2}} \\
\times \tf(\partial_{\eta_1}-\partial_{\eta_2}) 
\frac{e^{-\gamma_E {\eta_1}}}{\Gamma({\eta_1}+1)}\,
\frac{e^{-\gamma_E {\eta_2}}}{\Gamma({\eta_2}+1)}\,,
\end{multline}
where $\eta_1 =\eta_2= 2 A_{\Gamma} (\mu_j, \mu_s)$.  
In contrast to thrust, for heavy jet mass the full functional form of $\tf(\rL)$ is needed. For N${}^3$LL precision, we
need to know the hemisphere soft function, and hence $\tf(\rL)$ to two-loop order (NLO). Actually, to this
order, we only need one projection of the hemisphere soft function. For three-loop matching (NNLO),
we need an additional projection. These projections will be discussed in the next section.

One interesting feature of the hemisphere mass distribution is that the soft interference effects in $\tf(\rL)$ are
only relevant at $\alpha_s^2$, which is appropriate for N${}^3$LL resummation. Up to NNLL accuracy, the doubly differential
distribution is simply the product of
the  mass distributions in the two hemispheres. 
Explicitly,
\begin{equation}
  R(M_L^2,M_R^2) = \int_0^{M_L^2} \rd M_L^2{}' \int_0^{M_R^2}\rd M_R^2{}' \frac{\rd^2\sigma}{\rd M_L^2{}' \rd M_R^2{}'} = K(M_L^2) K(M_R^2)\,,
\end{equation}
where
\begin{multline}
  K(M^2) = 
\exp\left[ 2S(\mu_h,\mu_j)+2S(\mu_s,\mu_j)-A_H(\mu_h,\mu_s)+2A_J(\mu_j,\mu_s)\right]
 \left(\frac{Q^2}{\mu_h^2}\right)^{-A_\Gamma(\mu_h,\mu_j)} 
\, \\
\times\sqrt{H(Q^2,\mu_h)\tf(0)}\,
\widetilde j\Big( \ln\frac{\mu_s Q}{\mu_j^2}+\partial_{\eta},\mu_j\Big)\,
\widetilde s_\mu(\partial_{\eta},\mu_s)
\left(\frac{M^2}{\mu_s Q}  \right)^{{\eta}}
\frac{e^{-\gamma_E {\eta}}}{\Gamma({\eta}+1)}\,,
\end{multline}
and $\eta= 2 A_\Gamma(\mu_j,\mu_s)$.
Since, for NNLL resummation, the hard and jet functions are only needed to ${\mathcal O}(\alpha_s)$, the square-roots above simply mean take
one half of the $\alpha_s$ pieces. The fact that the distribution splits up in
this way was observed at NLL level in~\cite{Schwartz:2007ib}, and
is essential to the traditional NLL resummation~\cite{Catani:1992ua}.
This simplified factorization suggests that it may be possible
to calculate observables involving many more jets with NNLL resummation without having to disentangle soft interference effects.
Note that this factorization does not guarantee that large logs of $M_L^2/M_R^2$ can be resummed. However, it is possible that
the calculation of observables with only one scale, such as the sum of many jet masses, or a maximal jet mass, 
will simplify with SCET.


\section{Hemisphere Soft Function and Comparison to Fixed Order \label{sec:soft}}
The hemisphere soft function has been studied briefly in~\cite{Fleming:2007qr,Schwartz:2007ib,Fleming:2007xt} and
more thoroughly in~\cite{Hoang:2008fs}. It is a function of two scales, $k_L$ and $k_R$
as well as the renormalization group scale $\mu$. If ${\red n_L^\mu}$ is the direction of the left hemisphere and $k^\mu_L$ is the 
sum of the momenta of all the soft radiation entering
this hemisphere, then $k_L$ is the component of $k_L^\mu$ backwards to ${\red n_L^\mu}$. 
That is $k_L = (k_L \cdt {\red n_L})$.
$k_R$ is defined analogously. The soft function can be factorized into a perturbative, partonic part,
 and non-perturbative contribution which
has support of order $\Lambda_\mathrm{QCD}$. For now we deal only with the perturbative part, discussing non-perturbative effects in Section~\ref{sec:NP}.

As we have noted, the soft function itself factorizes. 
\begin{equation}
  \widetilde{s}(\rLO,\rLT,\mu) = \widetilde{s}_\mu(\rLO,\mu) \widetilde{s}_\mu(\rLT,\mu)
  \tf(\rLO -\rLT)
\end{equation}
where $\widetilde{s}(\rLO,\rLT,\mu)$ is the Laplace transform of $S(k_L,k_R,\mu)$, as in Eq.~\eqref{lapdef}, and 
$\rLO = \ln(\mu {\red  \nu_L} e^{\gamma_E})$, $\rLT = \ln(\mu {\red \nu_R} e^{\gamma_E})$.
The function $\widetilde{s}_\mu(\rL,\mu)$ is completely fixed by RG invariance in terms of the hard and jet
anomalous dimensions. It can be calculated in perturbation theory by demanding Eq.~\eqref{hjmdist} be independent of $\mu$.
This gives
\begin{multline} \label{smuexp}
\widetilde{s}_\mu({\rL},\mu) 
= \exp\Big[
  \left(\frac{\alpha_s}{4\pi}\right)  \left(
    -\rL^2 \Gamma_0 + \rL \gamma^S_0
  \right) 
+ \left(\frac{\alpha_s}{4\pi}\right)^2
    \left(\frac{2}{3}\rL^3 \beta_0 \Gamma_0 + \rL^2(-\Gamma_1 - \beta_0 \gamma_0^S) + \rL ( \gamma^S_1)
  \right)\\
+  \left(\frac{\alpha_s}{4\pi}\right)^3  \left(
    -\frac{2}{3} \rL^4 \beta_0^2 \Gamma_0 + \frac{2}{3} \rL^3 (\beta_1 \Gamma_0 + 2 \beta_0 \Gamma_1    + 2 \beta_0^2 \gamma_0^S)
    + \rL^2( - \Gamma_2 - \beta_1 \gamma_0^S - 2 \beta_0 \gamma_1^S)
    + \rL ( \gamma_2^S)
\right)\\
+\cdots \Big] \,.
\end{multline}
The $\mu$-independent part $\tf(\rL)$ must satisfy a number of constraints,
as discussed in~\cite{Hoang:2008fs}.

First of all, since the soft function is symmetric in
the two hemispheres, $\tf(\rL)$ must be an even function of $\rL$. Second of all, we know the function to order $\alpha_s$ by
explicit calculation. Writing
\begin{equation}
\tf(\rL) = 1+ \left(\frac{\alpha_s}{4\pi}\right)\tf{}_1(\rL) +\left(\frac{\alpha_s}{4\pi}\right)^2 \tf{}_2(\rL)+\cdots\, ,
\end{equation}
the one-loop result is that
\begin{equation} \label{softoneloop}
  \tf{}_1(\rL) =-C_F\pi^2 \,.
\end{equation}
The authors of~\cite{Hoang:2008fs} also observed that $\tf(\rL)$ is
constrained by the non-Abelian exponentiation theorem. Non-Abelian exponentiation implies constraints on powers of logarithms of $\mu$
in the full soft function. These constraints are satisfied by the explicit solution, since
 $\wt{s}_\mu(\rL,\mu)$ is an exponential. The theorem also restricts the $C_F^n$ color
structure in the soft function to be completely determined by the one-loop result, Eq.~\eqref{softoneloop}.
 Beyond this, however, $\tf(\rL)$
is unconstrained. It may even have more general dependence on $\rL$ than logarithms.
To determine $\tf(\rL)$, we must calculate the soft function perturbatively. The one-loop calculation has been done but  the
two-loop calculation, which is required for N${}^3$LL resummation, has not.

A simple alternative to calculating $\tf(\rL)$ at NNLO is to extract projections of $\tf(\rL)$ from numerical comparisons
to event shape calculations in full QCD. For example, thrust is only sensitive to $\tf(0)$. Writing
\begin{equation}
  \tf(0) = 1 + \left(\frac{\alpha_s}{4\pi}\right) \cone + \left(\frac{\alpha_s}{4\pi}\right)^2 \ctwo + \cdots \,,
\end{equation}
and comparing to Eq.~\eqref{softoneloop}, we see that $\cone = -C_F \pi^2$. The two-loop constant was determined numerically
in~\cite{Becher:2008cf} with the use of the {\sc event 2} program~\cite{Catani:1996jh}. The result is
\begin{equation} \label{c2s}
  \ctwo =(58 \pm 2) C_F^2 + (-60 \pm 1) C_F C_A + (43 \pm 1)C_F T_F n_f \quad \text{(Becher and Schwartz)} 
\end{equation}
This is in conflict with the prediction from non-Abelian exponentiation, which requires
the $C_F^2$ factor be $\frac{1}{2} \pi^4C_F^2 = 48.7C_F^2$.
The two-loop constant was also determined in~\cite{Hoang:2008fs}, using the same technique but imposing non-Abelian exponentiation. They found
\begin{equation}
  \ctwo =\frac{\pi^4}{2} C_F^2 +(-59\pm 2) C_F C_A + (44 \pm 3)C_F T_F n_f \quad \text{(Hoang and Kluth)} 
\end{equation}
The two results agree, except for the $C_F^2$ term. Indeed, the $C_F^2$ term seems to indicate a disagreement between the
numerical results of the {\sc event 2} program and the prediction from non-Abelian exponentiation. Since the
uncertainty in Eq~\eqref{c2s} is too small to explain this disagreement, it is reasonable also to expect the other 
color structures to be off. We should therefore allow for a systematic uncertainty on these fits in addition
to what is presented, which is essentially a statistical uncertainty associated with the fit. We discuss this more below.

Event shapes other than thrust are sensitive to the form of $\tf(\rL)$, not just $\tf(0)$.
This can be seen, for example, by the form of the heavy jet mass distribution in Eq.~\eqref{hjmdist}.
For N${}^3$LL resummation, the fixed order expansion is required to $\alpha_s^2$. The contribution
at this order involving $\tf(\rL)$ requires at most  $\tf{}_2(\rL)$, with the jet and hard functions at their tree-level values. Thus, the
required projection of the $\tf(\rL)$ for heavy jet mass is
\begin{align}
\ctwor &=
\tf{}_2(\partial_{\eta_1}-\partial_{\eta_2})
\left.\frac{e^{-\gamma_E {\eta_1}}}{\Gamma({\eta_1}+1)}\,
\frac{e^{-\gamma_E {\eta_2}}}{\Gamma({\eta_2}+1)}\right|_{\eta_1=\eta_2=0}
= \frac{1}{\pi}\int_0^{\pi}\tf{}_2(i \rL) \rd \rL \,.
\end{align}
The integral representation of $\ctwor$ 
is suggestive of a deeper relation between heavy jet mass and the hemisphere mass distribution,
however we do not have a physical explanation of why this particular moment appears. 
If $\tf(\rL)$ is a polynomial, this moment is very simple.
For example, if we assume
\begin{equation}\label{secondansatz}
  \tf(\rLM) = 1 + \left(\frac{\alpha_s}{4\pi}\right) \cone + \left(\frac{\alpha_s}{4\pi}\right)^2 
\left[\ctwo +\ctwoL\rL^2 +\ctwoQ \rL^4 \right] \,,
\end{equation}
then
\begin{equation} \label{c2qeq}
  \ctwor = \ctwo -\ctwoL \frac{\pi^2}{3} + \ctwoQ\frac{\pi^4}{5} \,.
\end{equation}
At NLO, the singular part of the heavy jet mass distribution only depends on $\tf(\rL)$ through $\ctwor$. Thus,
we can fit $\ctwor$ numerically the same way $\ctwo$ is fit with thrust.

\begin{figure}[t!]
\begin{center}
\psfrag{x}{$\rho$}
\psfrag{yyyyyyyy}{$\frac{1}{\sigma_0} \Delta\left[\rho \frac{\rd \sigma}{\rd \rho}\right]$}
\psfrag{yyyyyyyz}{}
\psfrag{CF}{$C_F$}
\psfrag{CA}{$C_A$}
\psfrag{NF}{$n_F$}
\psfrag{total}{$\text{total}$}
\includegraphics[width=0.9\textwidth]{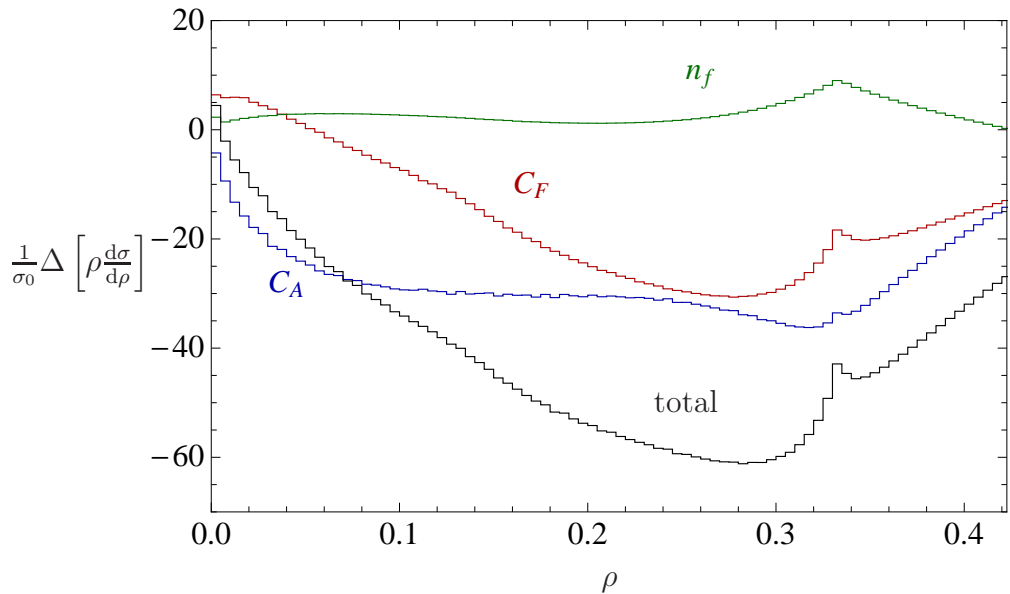}
\end{center}
\vspace*{-0.5cm}
\caption{A comparison of the full fixed-order calculations and expanded SCET at NLO.
Shown is the difference $\frac{1}{\sigma_0}\rho \Delta \frac{\rd \sigma}{\rd \rho}
=\rho( B(\rho) - D_B(\rho))$, where $B(\rho)$ is the full NLO $B$-function,
calculated with {\sc event 2} and $D_B(\rho)$ is the singular part,
calculated with SCET. The differences are separated by color structure, with the sum also shown.
The kink at $\rho=\frac{1}{3}$ is the maximum heavy jet mass for a  3-particle final state.
(See also Figure~\ref{fig:nlodiffcut12}.)
}
\label{fig:nlodiff}
\end{figure}

\begin{figure}[t!]
\begin{center}
\psfrag{y}[]{$\ctwor$}
\psfrag{x}[]{$\rmin$}
\includegraphics[width=0.7\textwidth]{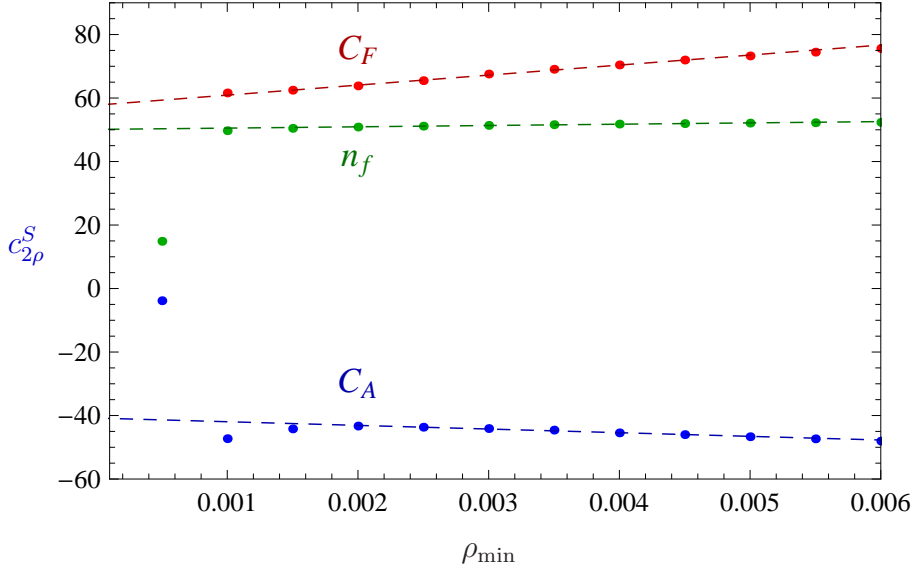}
\end{center}
\vspace*{-0.5cm}
\caption{Extraction of the two-loop constants in the soft function. The points correspond to the value of a lower bound
$\rmin$ applied to the fixed-order calculation. The lines are
interpolations among the points from $\rmin=0.002$ to $\rmin=0.005$ 
extrapolated to $\rho=0$ to extract
the constants.(See also Figure~\ref{fig:softNLOcut12}.)
}
\label{fig:softNLO}
\end{figure}

To determine $\ctwor$ we use the same technique used in~\cite{Becher:2008cf} for thrust, and in~\cite{Hoang:2008fs}
for a one-parameter family of event shapes. The basic idea is that the singular part of the heavy jet mass distribution
is known analytically, through SCET. The difference between the exact NLO heavy jet mass distribution and
this singular part is finite and can be integrated numerically.  This integral is then the total cross section at NLO minus the integral
of the singular part which is calculable analytically and depends on the constants $\ctwo$ for thrust and $\ctwor$ for heavy jet mass.

In more detail, the exact heavy jet mass distribution can be expanded as a series in $\alpha_s$
\begin{equation}
  \frac{1}{\sigma_0} \frac{\rd \sigma}{\rd \rho} =
  \left( \frac{\alpha_s}{2 \pi} \right) A(\rho)  + \left(
  \frac{\alpha_s}{2 \pi} \right)^2 B (\rho)  + \left(
  \frac{\alpha_s}{2 \pi} \right)^3 C (\rho) + \cdots\,.
\end{equation}
Each term in this series is singular at $\rho=0$. The singular parts can be written as a sum of distributions
\begin{equation}
  \frac{1}{\sigma_0} \frac{\rd \sigma}{\rd \rho} = \delta (\rho) D_{\delta} +
  \left( \frac{\alpha_s}{2 \pi} \right) \left[ D_A (\rho) \right]_+ + \left(
  \frac{\alpha_s}{2 \pi} \right)^2 \left[ D_B (\rho) \right]_+ + \left(
  \frac{\alpha_s}{2 \pi} \right)^3 \left[ D_C (\rho) \right]_+ + \cdots\,.
  \label{foscet}
\end{equation}
The functions $D_\delta$, $D_A(\rho)$, $D_B(\rho)$, and $D_C(\rho)$ are calculable in
SCET and we give them in Appendix~\ref{app:DABC}. 
Up to order $\alpha_s^2$, the only dependence on the unknown soft function
coefficient $\ctwor$ is in $D_\delta$, thus the shape of the singular part of the
NLO distribution is known completely.
The corresponding exact distributions in perturbative
QCD have been calculated for $\rho>0$ analytically for the $A$ function, and numerically for the $B$ and $C$ functions.
Since SCET produces the entire singular part of the distributions, the combination
\begin{equation} \label{diffdef}
 \frac{1}{\sigma_0} \Delta\left[ \rho \frac{\rd \sigma}{\rd \rho} \right]  = \rho B(\rho) - \rho D_B(\rho) \,,
\end{equation}
should vanish at $\rho=0$. We show this difference separated by color structure in Figure~\ref{fig:nlodiff}. The $B$ functions
are calculated using the Monte Carlo program  {\sc event 2}~\cite{Catani:1996jh} with $10^{10}$ events.
Curiously, while the $C_F$ and $C_A$ color structures do not seem to go to $0$ as $\rho\to 0$, their sum does.

\begin{figure}[t!]
\begin{center}
\psfrag{N2}[l]{\small $N^2$}
\psfrag{N0}[l]{\small $N^0$}
\psfrag{Nm}[l]{\small $1/N^{2}$}
\psfrag{Nnf}[l]{\small $n_f N$}
\psfrag{NfNm}[l]{\small $\phantom{a}n_f/N$}
\psfrag{Nf2}[l]{\small $n_f^{2}$}
\psfrag{sig}[t]{ $\frac{10^{-3}}{\sigma_0} \rho \frac{\rd\sigma}{\rd \rho}$}
\psfrag{x}[b]{\small $\phantom{abc}-\log\rho$}
\includegraphics[width=\textwidth]{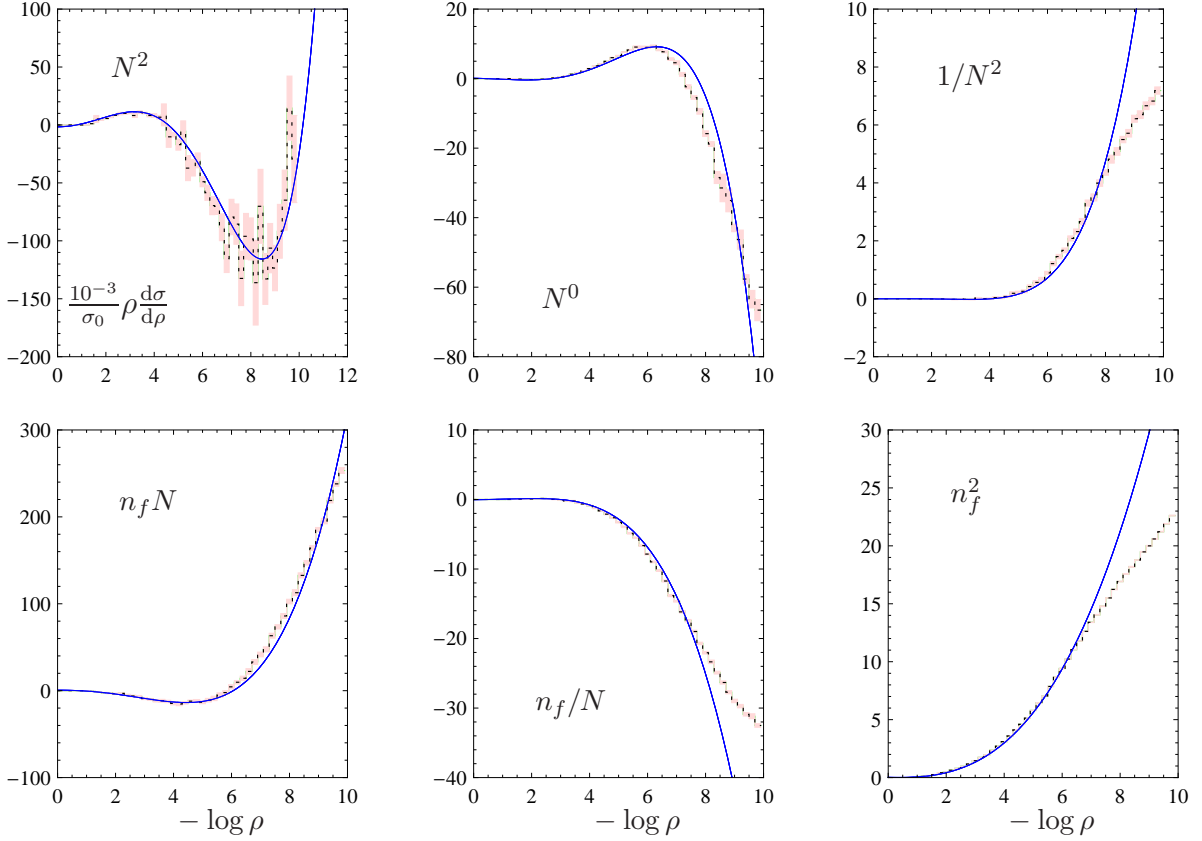}
\end{center}
\vspace*{-0.5cm}
\caption{\label{fig:NNLOcolslog}
Comparison of the full NNLO heavy jet mass distribution $C(\rho)$ (dashed black histograms)~\cite{GehrmannDeRidder:2007jk}
 and the singular terms $D_C(\rho)$ (blue curves).
The light-red areas are an estimate of the statistical uncertainty. The 
uncertainties on $\ctwo$ and $\ctwoL$, in Eq~\eqref{ourchoice}, are not visible.
The disagreement at very small $\rho$ is due to
the infrared cut-off of $y_0=10^{-5}$ for the NNLO calculation.
 It is expected that the agreement would
improve if this cutoff were lowered, as can be seen in the analogous thrust plot in~\cite{Becher:2008cf}.
(See also Figure~\ref{fig:NNLOcolslogcut7}.)
}
\end{figure}

\begin{figure}[t!]
\begin{center}
\psfrag{N2}[l]{\small $N^2$}
\psfrag{N0}[l]{\small $N^0$}
\psfrag{Nm}[l]{\small $1/N^{2}$}
\psfrag{Nnf}[l]{\small $n_f N$}
\psfrag{NfNm}[l]{\small $n_f/N$}
\psfrag{Nf2}[l]{\small $n_f^{2}$}
\psfrag{sig}[t]{$\frac{1}{\sigma_0} \Delta\left[\rho \frac{\rd \sigma}{\rd \rho}\right]$}
\psfrag{x}[b]{\small $\rho$}
\includegraphics[width=\textwidth]{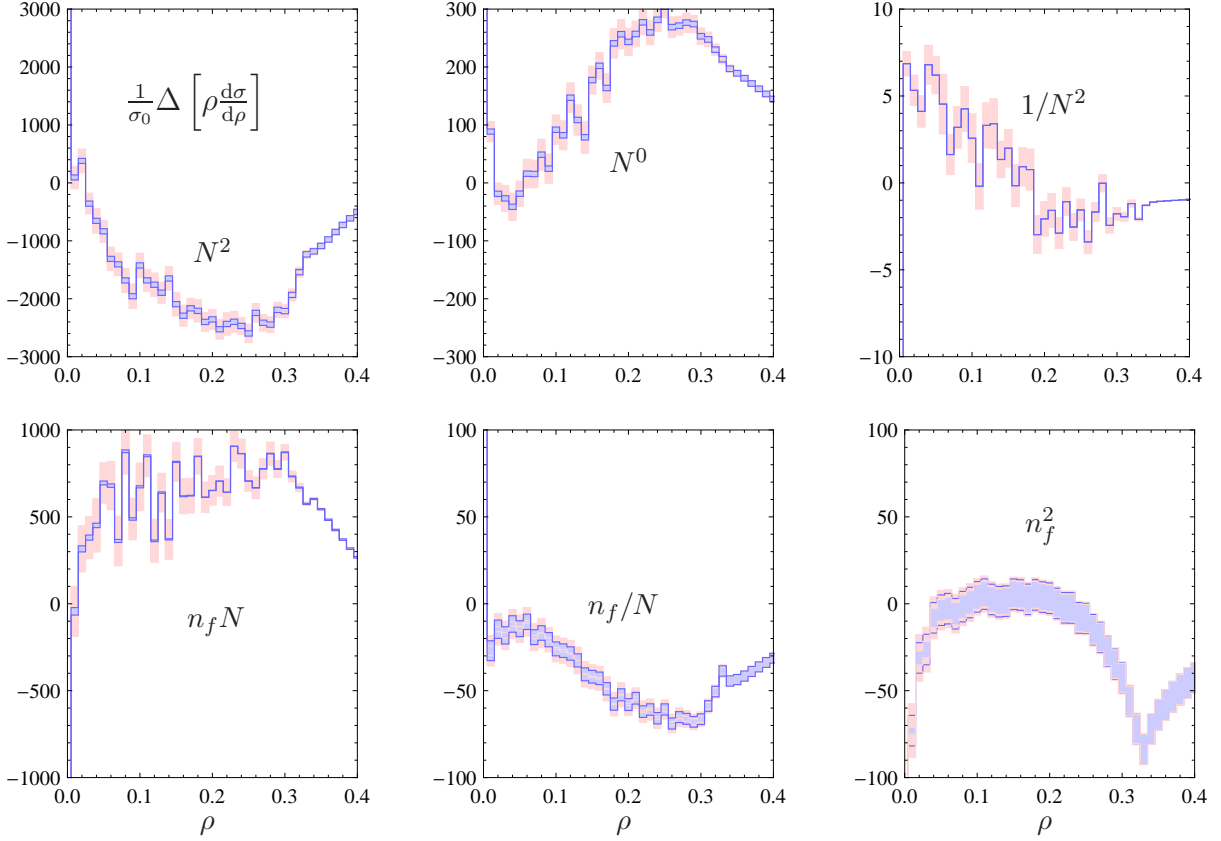}
\end{center}
\caption{\label{fig:nnlodiff} 
The difference, $\frac{1}{\sigma_0}\rho \Delta \frac{\rd \sigma}{\rd \rho}
=\rho( C(\rho) - D_C(\rho))$, between the full NNLO heavy jet mass distribution and the singular terms.
The light-red areas are an estimate of the statistical uncertainty, from~\cite{GehrmannDeRidder:2007jk}. The blue
band is the uncertainty due to $\ctwo$ and $\ctwoL$, in Eq~\eqref{ourchoice}.
These curves should all go to 0 at $\rho=0$.  The $C(\rho)$ distributions are all calculated with an infrared cutoff
of $y_0 = 10^{-5}$. (See Figure~\ref{fig:nnlodiffcut7} for the same figure with $y_0=10^{-7}$.)
}
\end{figure}

With these difference functions, it is straightforward to extract $\ctwor$ as in~\cite{Becher:2008cf} and~\cite{Hoang:2008fs}. Although
the difference $\rho B(\rho) - \rho D_B(\rho)$ is regular at $\rho=0$, the two functions are separately divergent. Since  $D_B(\rho)$ is
only known numerically, the difference is numerically unstable at small $\rho$.  To do the $\ctwor$ extraction,
we use the same procedure as in~\cite{Becher:2008cf} and impose an lower bound and take the limit that the bound is removed.
The extracted values as a function of this lower bound $\rmin$ are shown in Figure~\ref{fig:softNLO}. We then fit a line
in the region $0.002 \le \rmin \le 0.005$ and extrapolate to $\rmin=0$.
The result is
\begin{equation}
   \ctwor =(58 \pm 2) C_F^2 + (-41 \pm 2) C_F C_A + (50 \pm 1)C_F T_F n_f \,.
\end{equation}
Note that the $C_F$ and $C_A$ curves have problems at small $\rmin$, in agreement with what is seen in Figure~\ref{fig:nlodiff}.
Since the approach is linear up to around $\rmin\sim 0.002$, it is likely 
that this divergence is an unphysical systematic problem with the Monte Carlo,
and not due to low statistics or a discrepancy with theory.~\footnote{We thank A. Hoang for a discussion of this point.}
As with $\ctwo$, non-Abelian exponentiation implies that the $C_F^2$ term should be $\frac{1}{2}\pi^4 \approx 48.7$. Thus,
our uncertainty of $\ctwor$ from the extrapolation is probably too small and we will therefore inflate the errors
by a factor of 5. (See also Eq.~\eqref{ctwornew} and Figures~\ref{fig:nlodiffcut12} and~\ref{fig:softNLOcut12} for
an update.)

To calculate the heavy jet mass distribution to N${}^3$LL+NNLO accuracy, we must match to the NNLO fixed
order distribution. This requires the singular parts of heavy jet mass to $\alpha_s^3$, that is, the function
$D_C(\rho)$ in Eq.~\eqref{foscet}.
To derive this, we do not need the finite part of the soft function at $\alpha_s^3$, $\tf{}_3(\rL)$, since this piece
 only contributes to the $\alpha_s^3$ part of $D_\delta$, which is not required for matching. We do,
however, need another projection of the  $\alpha_s^2$ soft function, of the form 
\begin{align} \label{ctwozdef}
\ctwoz & = (\partial_{\eta_1}+\partial_{\eta_2})
\tf{}_2(\partial_{\eta_1}-\partial_{\eta_2})
\left.\frac{e^{-\gamma_E {\eta_1}}}{\Gamma({\eta_1}+1)}\,
\frac{e^{-\gamma_E {\eta_2}}}{\Gamma({\eta_2}+1)}\right|_{\eta_1=\eta_2=0}\\
&=\frac{2}{\pi} \int_0^\pi \tf{}_2(i \rL) \ln\left[ 2 \cos(\frac{\rL}{2}) \right]\rd\rL \,.
\end{align}
Again, we have no physical explanation of the intriguing integral definition in the second line.
This projection also simplifies with a polynomial soft function. For example, with Eq~\eqref{secondansatz}
\begin{equation}
  \ctwoz = 4\zeta_3\ctwoL +( -8\pi^2\zeta_3+ 48\zeta_5) \ctwoQ \,.
\end{equation}
The prediction from SCET for $D_C(\rho)$ with its explicit dependence
only on $\ctwor$ and $\ctwoz$ is given in Appendix~\ref{app:DABC}.
There are only three color structures which depend on $\ctwoz$ at all.

In order to extract the $\rL$ dependence of $\tf{}_2(\rL)$, we could attempt to fit $\ctwoz$
with the shapes of the NNLO distributions. An alternative, as pursued by Hoang and Kluth in~\cite{Hoang:2008fs}, is
to use the other event shapes beyond thrust and heavy jet mass at NLO. These authors considered a weighted
sum of the jet masses, $\tau_\alpha = \frac{2}{1+\alpha}(\alpha M_L^2 + M_R^2)/Q^2$. This form
leads to a singular distribution which depends on $\tf{}_2(\ln \alpha)$, hence combining event shapes with different $\alpha$
can probe the entire function $\tf{}_2(\rL)$.
Their fits show good agreement with the form
\begin{equation}~\label{softansatz}
  \tf(\rLM) =  1 + \left(\frac{\alpha_s}{4\pi}\right) \cone + \left(\frac{\alpha_s}{4\pi}\right)^2 
\left[\ctwo +\ctwoL\rLM^2\right] \,,
\end{equation}
which they have argued is likely to be the exact 2-loop soft function.
We will therefore assume this
form of the soft function as well, in order to proceed with the N${}^3$LL+NNLO $\alpha_s$ fits.~\footnote{
There is a subtlety about these $\tau_\alpha$ event shapes because of non-global logarithms~\cite{Dasgupta:2001sh}. 
For example, for very large or small $\alpha$, these
event shapes reduce to the left or right hemisphere mass, which are known to have non-global logs. 
Since $\tau_\alpha\to0$ forces the massless dijet threshold, in which the SCET factorization theorem is derived,
only up to corrections of order $\ln\alpha$,  it is not completely clear
that SCET will reproduce all of the $\alpha$-dependence of the singular terms in $\tau_\alpha$.}

With this soft function and the thrust fit values in Eq.~\eqref{c2s},
 our fit for $\ctwor$ translates into a fit for $\ctwoL$ (cf. Eq.\eqref{c2qeq} with $\ctwoQ=0$)
The result is
\begin{equation}
  \ctwoL =(0 \pm 2) C_F^2 + (-5.8 \pm 1.5) C_F C_A + (-2.2 \pm 1)C_F T_F n_f \,.
\end{equation}
Using a similar technique, but imposing the constraint from non-Abelian exponentiation, Hoang and Kluth found
results consistent with ours
\begin{equation}
  \ctwoL =(0) C_F^2 + (-6.5 \pm 2) C_F C_A + (1.3 \pm 2)C_F T_F n_f \quad \text{(Hoang and Kluth)} 
\end{equation}
Note that for $\ctwoL$, the $C_F^2$ coefficient comes out
to be consistent with the prediction from non-Abelian exponentiation. Since $\ctwoL$ comes from the difference
between the values extracted from thrust and the values extracted from heavy jet mass, the systematic problem with
{\sc event 2} may be cancelling in the difference. Thus, we will inflate our uncertainties on $\ctwoL$ by only a factor of $2$.

In summary, for the rest of this paper, we will take
\begin{align} \label{ourchoice}
  \tf(\rLM) &=  1 + \left(\frac{\alpha_s}{4\pi}\right) \cone + \left(\frac{\alpha_s}{4\pi}\right)^2 
\left[\ctwo +\ctwoL\rLM^2\right]\\
  \ctwo  &= \frac{\pi^4}{2} C_F^2 + (-60 \pm 10)  C_F C_A + (43 \pm 5)C_F T_F n_f\\
  \ctwoL &= (0) C_F^2 + (-6 \pm 3) C_F C_A + (-2 \pm 2)C_F T_F n_f \, ,
\end{align}
so that
\begin{align} \label{c2def}
 \ctwoz &=4 \zeta_3 \ctwoL \quad \text{and}\quad \ctwor = \ctwo - \frac{\pi^2}{3} \ctwoL
 \,.
\end{align}
The uncertainty on $\alpha_s$ due to the uncertainty on these numbers will be included in the fits.

Before moving on the $\alpha_s$ extraction, we can compare the SCET prediction for the singular parts of the NNLO
distribution to the exact results, as was done for thrust in~\cite{Becher:2008cf}.
To do this, we use $D_C(\rho)$ from Appendix~\ref{app:DABC} with the substitutions in Eq.~\eqref{c2def}.
This lets us compare to the $C$ functions
in the NNLO distribution, from~\cite{GehrmannDeRidder:2007jk}.
Plots of $\rho D_C$ and $\rho C$ are shown in Figure~\ref{fig:NNLOcolslog} as functions
of $\log \rho$. The uncertainty on $\ctwo$ and $\ctwoL$ is included, but not visible in these plots.
Although the agreement is not perfect at very small $\rho$, it is expected to improve,
as we seen for thrust in~\cite{Becher:2008cf}, as
the the infrared cutoff used in the NNLO calculation is reduced from the value $y_0 = 10^{-5}$ used
here. A version of this plot with cutoff $y_0 = 10^{-7}$ has been included as Figure~\ref{fig:NNLOcolslogcut7},
confirming our expectations.

The difference between the full NNLO distribution and its singular parts,
as in Eq.~\eqref{diffdef}, is shown in Figure~\ref{fig:nnlodiff}. These curves,
for all color structures, should go to zero at $\rho=0$. For most of the color structures, this looks plausible,
although the $1/N^2$ color structure, corresponding to the $\alpha_s^3 C_F^3$ coefficient in the heavy jet mass distribution
which is fixed by non-Abelian exponentiation,
looks a bit suspicious. 
 Because this constant is known, we have not included an associated uncertainty.
The discrepancy is likely due to the infrared cutoff $y_0=10^{-5}$ used for these plots (an update with
 $y_0 = 10^{-7}$ is included as Figure~\ref{fig:nnlodiffcut7}).
Note that even if the Ansatz
in Eq.~\eqref{ourchoice} were wrong, a general dependence on $\ctwoz$ will only affect some of the
color structures, and even then would only generate an overall up or down shift in these curves (cf. the form
of $D_C$ in Eq.~\eqref{DCform}).


\section{$\alpha_s$ extraction and error analysis \label{sec:fit}}
In the previous section, we  determined the unknown coefficients in the hemisphere soft function and checked the singular
terms against the exact NLO and NNLO heavy jet mass distributions. Now we are ready to compare to data and
fit for the strong coupling constant $\alpha_s$.  The procedure we follow is identical to the procedure used
for thrust in~\cite{Becher:2008cf}, so we refer the reader to that paper for missing details.

For heavy jet mass, as for thrust, we match to the fixed order distribution via
\begin{equation}
 \frac{1}{\sigma_0}  \frac{\rd \sigma}{\rd \rho} = \frac{1}{\sigma_0}
  \frac{\rd \sigma_2}{\rd \rho}  + r(\rho) \,,
\end{equation}
with
\begin{equation}
 r(\rho)
 = \left( \frac{\alpha_s}{2
  \pi} \right) \left[ A (\rho) - D_A (\rho) \right]
  + \left( \frac{\alpha_s}{2 \pi} \right)^2 \left[ B (\rho) - D_B (\rho)
  \right] + \left( \frac{\alpha_s}{2 \pi} \right)^3 \left[ C (\rho) - D_C
  (\rho) \right]\label{scet3} \,,
\end{equation}
and $D_A, D_B$ and $D_C$ are given in Appendix~\ref{app:DABC}. The $A$ function
is known analytically, and is the same as for thrust (see~\cite{Schwartz:2007ib}). For $B(\rho)$ we use the output
of {\sc event 2}~\cite{Catani:1996jh}, and for $C(\rho)$ we use the NNLO calculation which has been provided to us
by the authors of~\cite{GehrmannDeRidder:2007hr}.
We normalize to the total
hadronic cross section at order $\alpha_s^2$, which is
\begin{equation}
  \frac{\sigma_{\rm had}}{\sigma_0} = 1 + \frac{\alpha_s}{4 \pi} \left[ 3 C_F \right]
  + \left( \frac{\alpha_s}{4 \pi} \right)^2 \left[  C_F C_A \left(
  \frac{123}{2} - 44 \zeta_3 \right) +  C_F T_F n_f \left( - 22 + 16
  \zeta_3 \right) - C_F^2 \frac{3}{2} \right]\, . \label{sigtot}
\end{equation}
Since the data is binned, what we actually use for the theory prediction is the
difference between the integrated heavy jet mass distribution evaluated at the
bin edges: $R_\rho(\rho_2)-R_\rho(\rho_1)$. Our fit ranges are chosen to be the same as in~\cite{Dissertori:2007xa},
so that we can use their values for the systematic experimental uncertainties.

\begin{figure}[t]
\psfrag{y}[l]{$\!\!\!\!\!\frac{\mathrm{fit-data}}{\mathrm{data}}$}
\psfrag{a1}[l]{\small $\alpha_s(m_Z) = 0.1214$}
\psfrag{a2}[l]{\small $\alpha_s(m_Z) = 0.1168$}
\psfrag{x}[l]{$\rho$}
\begin{center} 
\begin{tabular}{ll}
\hspace{-0.3cm}
\includegraphics[width=0.50\textwidth]{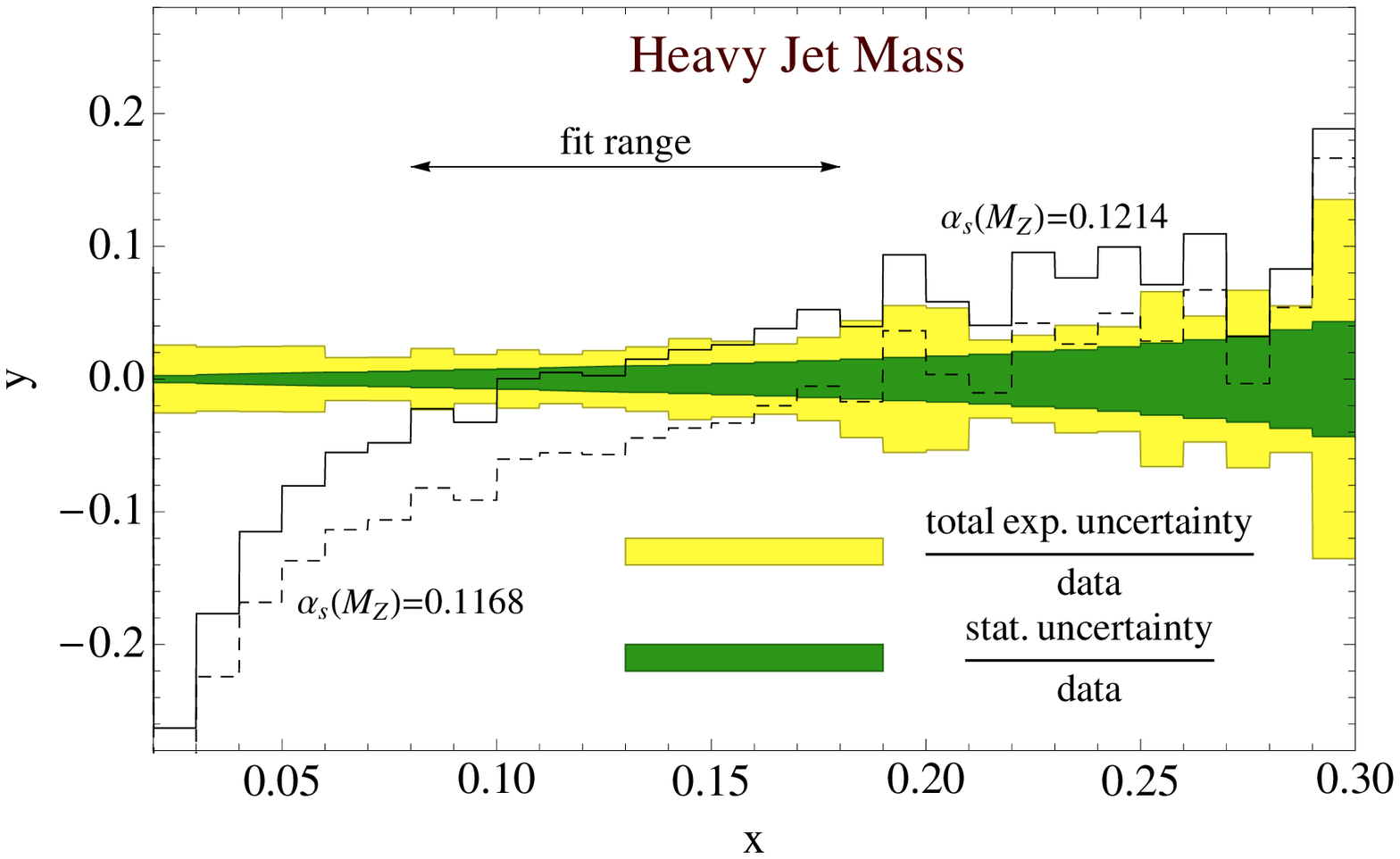} &
\psfrag{x}[l]{$\tau$}
\hspace{-0.4cm}
 \includegraphics[width=0.50\textwidth]{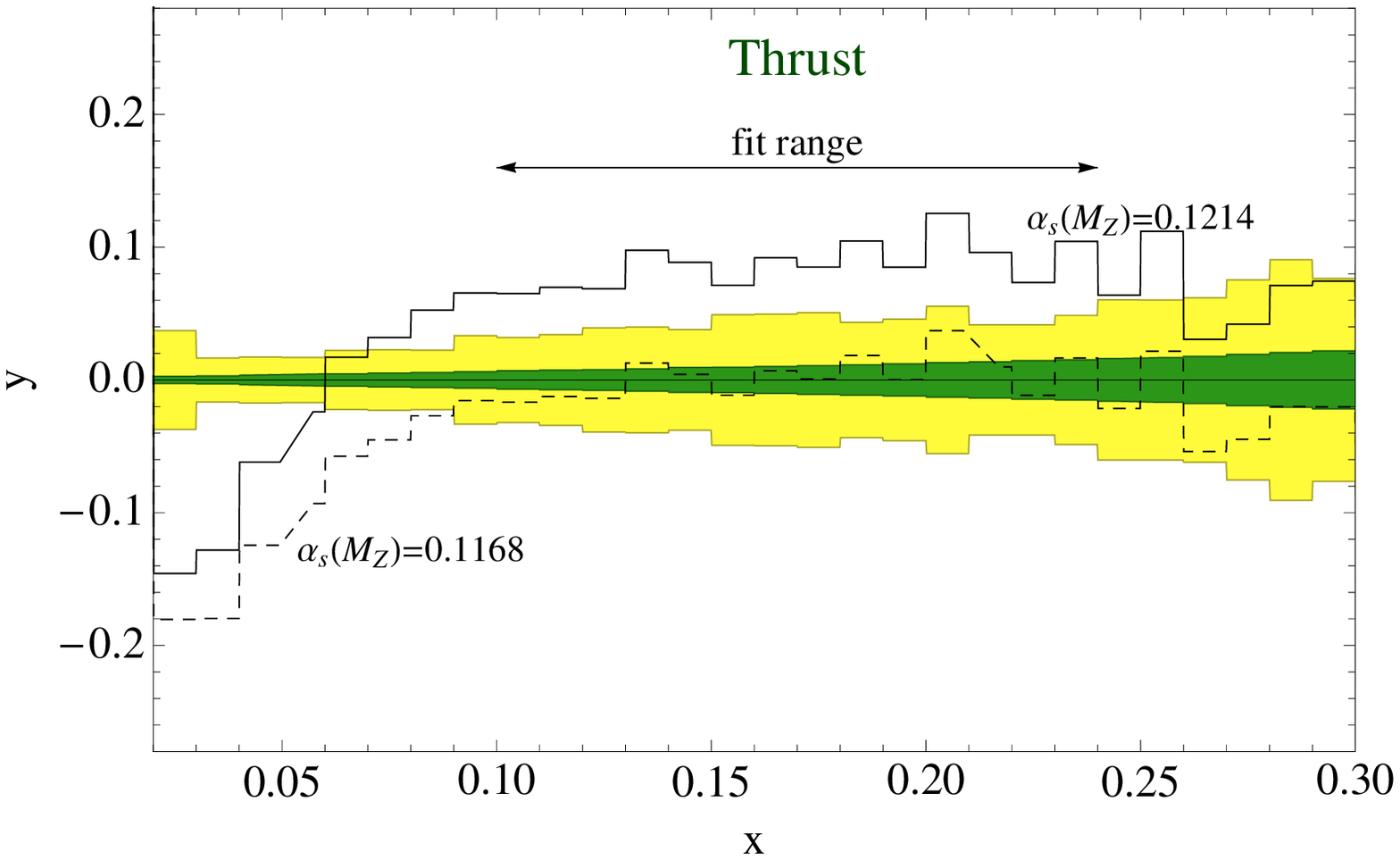} \\
\end{tabular}
\end{center}
\vspace*{-0.4cm}
\caption{Relative error for best fit to {\sc aleph} data at $91.2$ GeV. 
The inner green band includes only
statistical uncertainty, while the outer yellow band includes statistical, systematic and generator
uncertainties. 
The solid black line is for $\alpha_s(M_Z) = 0.1214$, the best fit value for heavy jet mass,
while the dashed line has $\alpha_s(m_Z) = 0.1168$, the best fit for thrust. 
The fit ranges, $0.08 < \rho < 0.18$ and $0.1 < \tau < 0.24$,
are taken  from~\cite{Dissertori:2007xa}.}
\label{fig:chis}
\end{figure}

The left panel of Figure~\ref{fig:chis} shows a comparison of the theory prediction for heavy jet mass
to the {\sc aleph} data at 91.2 GeV. 
These curves use the default scale choices
\begin{equation}
  \mu_h = Q,\quad \mu_j = Q \sqrt{\rho},\quad \mu_s = Q \rho \,.
\end{equation}
These scales are the natural ones to minimize the large logarithms, and
can be read off the formula in Eq.~\eqref{hjmdist}.
The best fit value of $\alpha_s$ for heavy jet mass is $\alpha_s(m_Z)=0.1214$. We show also in
the same figure, the heavy jet mass distribution for $\alpha_s(m_Z) = 0.1168$,
which is the value of $\alpha_s$ derived in~\cite{Becher:2008cf} from the fit to the thrust distribution at the same energy.
In the right panel of Figure~\ref{fig:chis}, we show a comparison to data for thrust, with the same values of $\alpha_s$.
Overall, the fit to thrust is a much better fit. For heavy jet mass, the best fit gives $\chi^2$/d.o.f.=67/9 using statistical
uncertainties only, while for thrust,   $\chi^2$/d.o.f.=32.5/13. The
relatively poor fit for heavy jet mass can be plainly seen in the figure. For thrust, the relative distribution 
is flat over the fit range (dashed curve, right panel), while for heavy jet mass, it is increasing (solid curve, left panel).
This coordinates with the relatively larger power corrections that we will find in the next section.

\begin{figure}[t!]
\psfrag{x}[]{$\rho$}
\psfrag{y}[]{$\frac{1}{\sigma}\frac{\rd \sigma}{\rd \rho}$\:\:\:}
 \begin{center}
 \includegraphics[width=\textwidth]{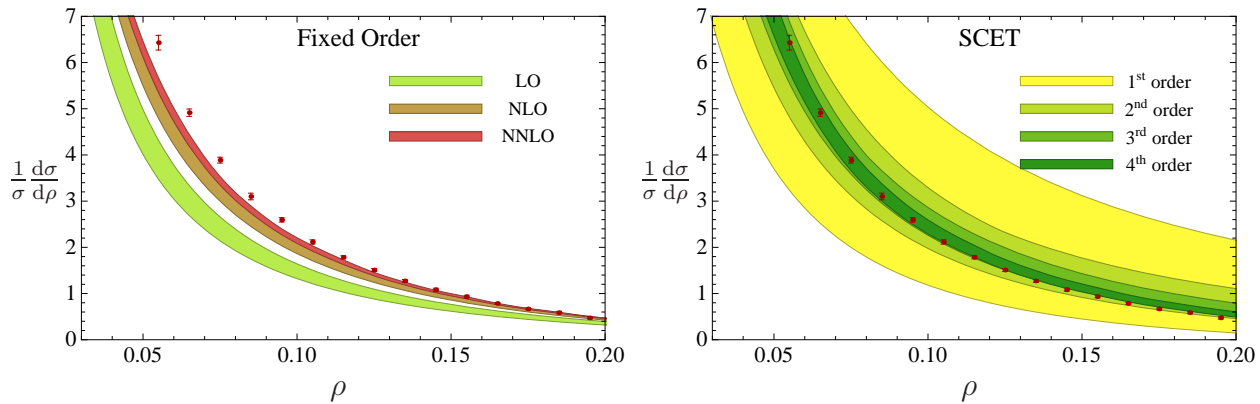}
 \end{center}
 \vspace{-0.8cm}
\caption{Convergence of resummed and fixed-order distributions. {\sc{aleph}} data (red)
at $91.2$ GeV is included for reference. All plots have $\alpha_s(m_Z)=0.1214$.}
\label{fig:sfconv}
\end{figure}

\begin{figure}[t!]
\psfrag{y}[]{\small $\rho \frac{1}{\sigma}\frac{\rd \sigma}{\rd \rho}$}
\psfrag{x}[]{\tiny $\rho$}
\begin{center}
\begin{tabular}{ll}
\includegraphics[width=0.47\textwidth]{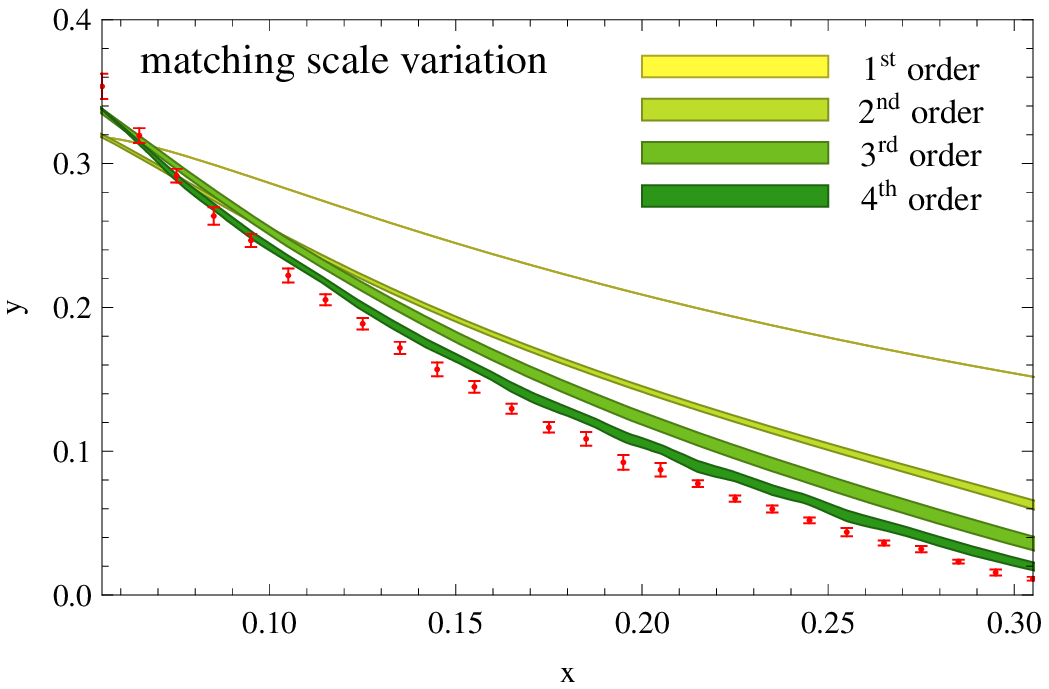} & \includegraphics[width=0.47\textwidth]{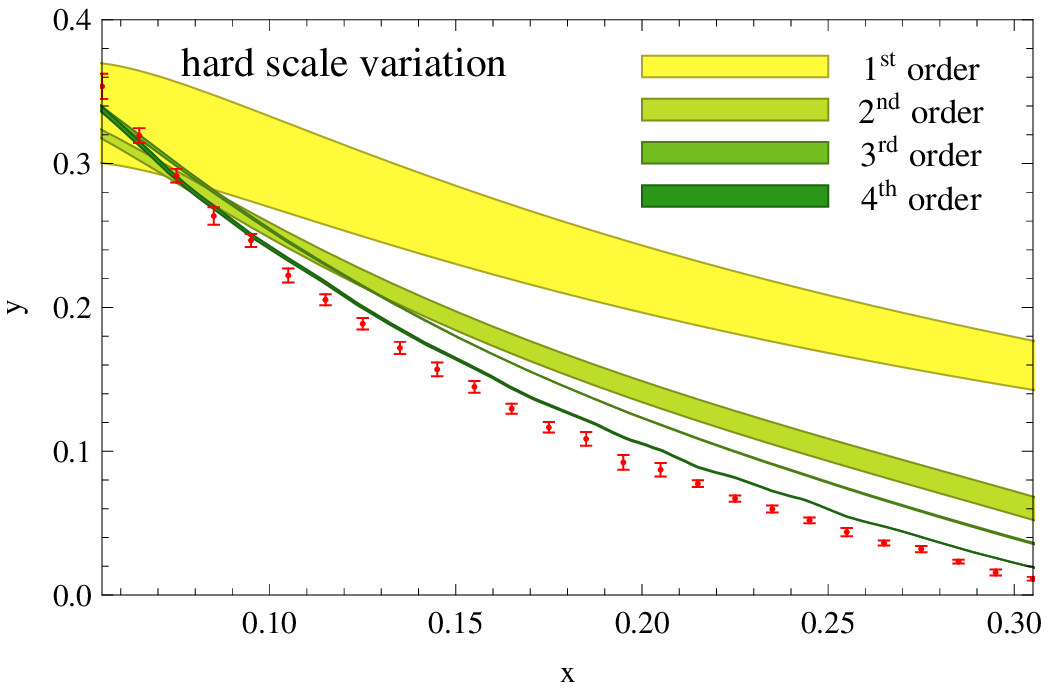} \\
\includegraphics[width=0.47\textwidth]{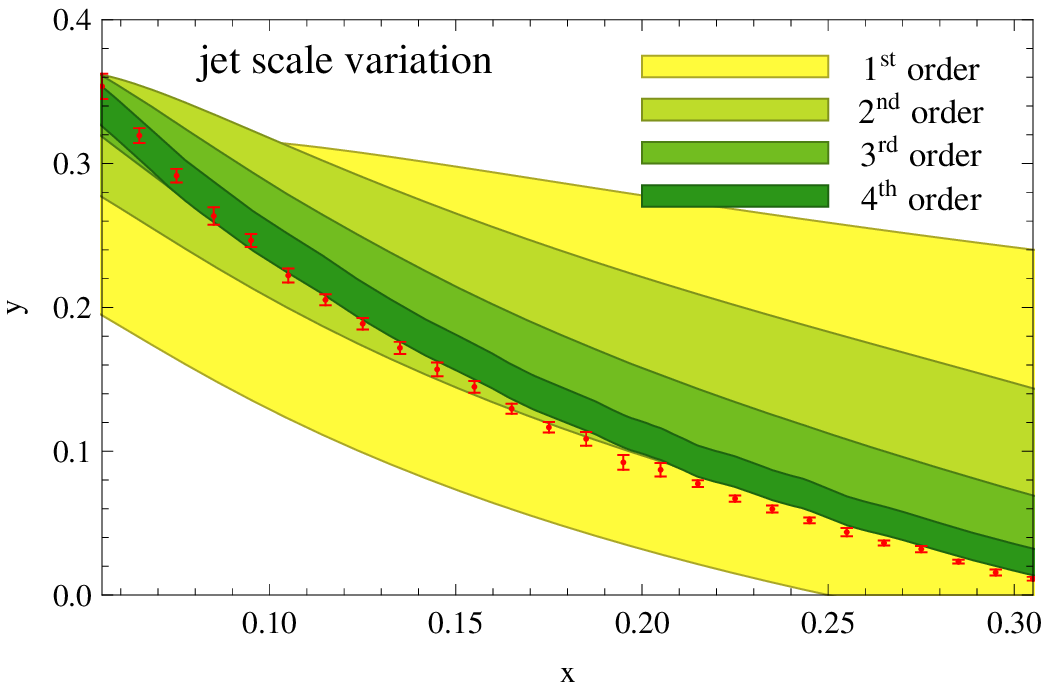} & \includegraphics[width=0.47\textwidth]{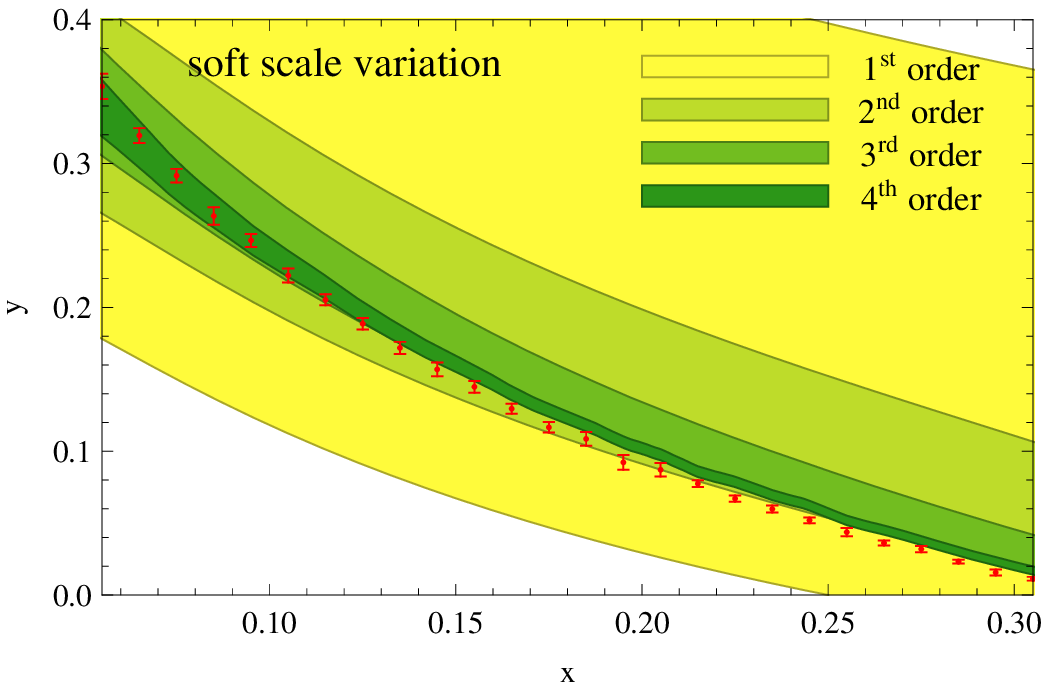}  \\
\includegraphics[width=0.47\textwidth]{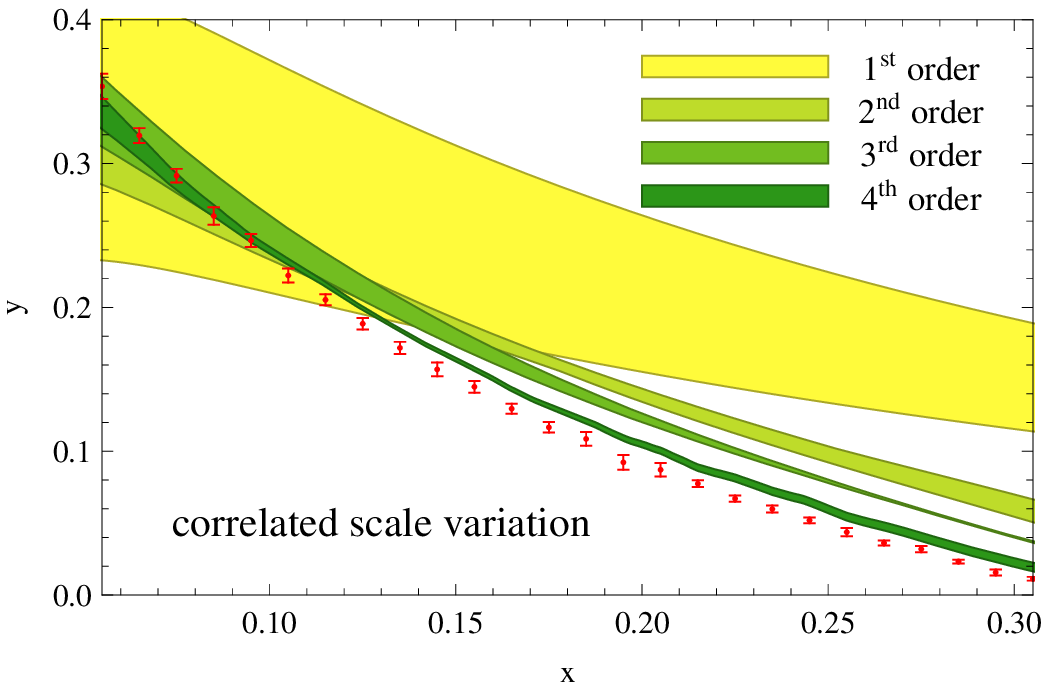} & \includegraphics[width=0.47\textwidth]{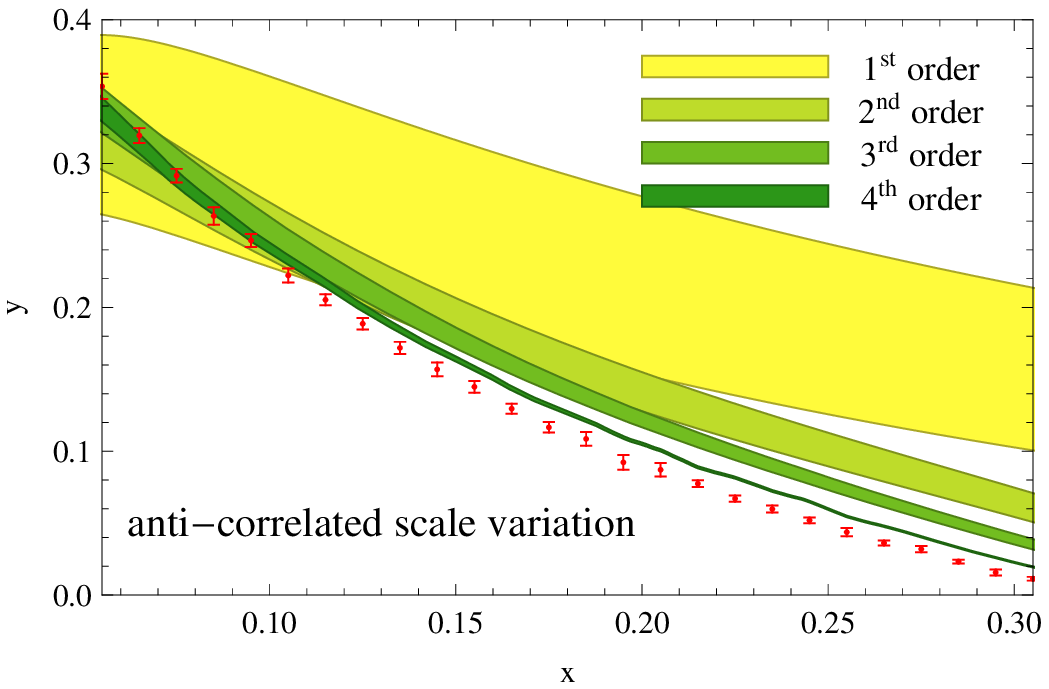}
\end{tabular}
\end{center}
\vspace{-0.8cm}
\caption{Perturbative uncertainty at $Q=91.2\,{\rm GeV}$. 
Each of the scales is varied separately by a factor of two around the default value. We show $\frac{1}{\sigma}\rho \frac{\rd \sigma}{\rd \rho}$
and, for reference, {\sc aleph} data at \lepone  scaled by the central value of each bin. 
All plots have $\alpha_s(m_Z)=0.1214$.}
\label{fig:MHJSBC}
\end{figure}
Next, we look at the uncertainties on the theoretical prediction. 
As with thrust, in~\cite{Becher:2008cf}, we consider
first separate variations of $\mu_h$, $\mu_j$, $\mu_s$ and the scale $\mu_m$ where the matching is done by factors
of $2$. Figure~\ref{fig:sfconv} shows the effect of the envelope of these variations on the heavy jet mass distribution, for
four orders in perturbation theory. We use the same definitions for the various orders as in~\cite{Becher:2008cf}:
\begin{center}
\begin{tabular}{|c|c|c|c|c|c|}
\hline 
Order & resum. &  $\Gamma_\text{cusp}$ & $\gamma_n$ & $c_n$ & matching\\
\hline
\first~order  & NLL & 2-loop & 1-loop & tree & -- \\ 
\second~order & NNLL & 3-loop & 2-loop & 1-loop & LO\\
\third~order & N${}^3$LL & 4-loop & 3-loop & 2-loop & NLO\\
\fourth~order & N${}^3$LL & 4-loop & 3-loop & 3-loop & NNLO\\
\hline
\end{tabular}
\end{center}
The first three orders correspond to traditional counting in renormalization-group improved perturbation theory,
while \fourth order simply uses all the available information.

Next, we consider, the separate variations.
The bands in the first four panels of Figure~\ref{fig:MHJSBC} show the effect of the  scale uncertainties. 
The bottom two panels of Figure~\ref{fig:MHJSBC} show the effect of the more natural correlated and
anti-correlated scale variations introduced in~\cite{Becher:2008cf}.
The correlated variation is defined to hold $\mu_j / \mu_s$ fixed. So we vary
\begin{equation}
  \mu_j \rightarrow c \sqrt{\tau} Q, \hspace{1em} \mu_s \rightarrow c \tau Q \,,
  \hspace{1em} \text{$\frac{1}{2} < c < 2$}\,.
\end{equation}
This probes the upper and lower limits on $\mu_j$ and $\mu_s$, but avoids the
unphysical region where $\mu_s < \mu_j$ or $\mu_h < \mu_j$. The orthogonal anti-correlated variation is 
defined to hold $\mu_j^2 / (Q \mu_s)$ fixed. It is
\begin{equation}
  \mu_j^2 \rightarrow a Q^2  \tau \hspace{1em} \mu_s \rightarrow a Q  \tau,
  \hspace{1em} \frac{1}{\sqrt{2}} < a < \sqrt{2}\,.
\end{equation}
This is independent from the correlated mode but again avoids unphysical scale choices.

Overall, we find good convergence order-by-order in perturbation theory. However, some of the 
higher-order scale variations are outside of the range of the lower orders. This was not the case for thrust,
where the central value of the prediction was much more stable. Nevertheless, for both thrust and
heavy jet mass, the complete perturbative uncertainty, defined as the envelope of the various variations
(that is, the maximum and minimum over them) does have the higher-order bands contained within the lower order bands,
as can be seen in Figure~\ref{fig:sfconv}.

Next, we fit the theoretical prediction to the {\sc aleph} data from 91.2 to 206 GeV~\cite{Heister:2003aj} and extract $\alpha_s$. The fit
is done by minimizing the $\chi^2$, using experimental statistical uncertainties, for the theory prediction with default scale choices. 
The {\bf statistical} error on $\alpha_s$ is determined by variations around this minimum. The {\bf perturbative}
uncertainty is extracted with the uncertainty band method~\cite{Jones:2003yv},
exactly as in~\cite{Becher:2008cf} for thrust. The envelope over the hard, matching, correlated and anti-correlated scale variations
are included in this extraction. We also include an additional {\bf soft} uncertainty associated with the errors
in the extraction of $\ctwo$ and $\ctwoL$. These are computed by fitting $\alpha_s$ within
the errors on $\ctwo$ and $\ctwoL$  in Eq.~\eqref{ourchoice}, and taking the difference with the central value as the uncertainty.
The soft and perturbative uncertainties are assumed uncorrelated.
The {\bf systematic} uncertainties are taken from~\cite{Dissertori:2007xa}. To
use these uncertainties, we are forced to keep our fit ranges the same as in~\cite{Dissertori:2007xa}.
The {\bf hadronization} uncertainties are also taken from~\cite{Dissertori:2007xa}, which are based on Monte Carlo
simulations. Note that, as in~\cite{Becher:2008cf}, we use the uncertainties from~\cite{Dissertori:2007xa} but
do not correct for hadronization. Hadronization will be discussed in detail in Section~\ref{sec:NP}.
 Finally,
the values for each energy are combined with a weight inversely proportional to the square of that
energy's total error. The statistical uncertainties are assumed uncorrelated, and combined in quadrature, while
for the other uncertainties a linear weighted average is performed. The results are tabulated in Table~\ref{tab:alephresults}.

\begin{table}
\begin{center}
  \begin{tabular}{|c|c|c|c|c|c|c|c|c|c|} 
    \hline
    Q    & 91.2 & 133  & 161  &  172 &  183 & 189  & 200  & 206  & AVG\\ \hline  \hline
\multirow{2}{*}{fit range}  
 &  0.08  & 0.06 & 0.06 & 0.06 & 0.06 & 0.04 & 0.04 & 0.04 & \multirow{2}{*}{--}\\
  & 0.18  & 0.25 & 0.25 & 0.25 & 0.25 & 0.20 & 0.20 & 0.20 & \\ 
\hline
       $\chi^2$/d.o.f.\ & 67/9 & 2.3/4 & 0.66/4 & 1.8/4 & 5.2/4 & 1.1/4 & 8.8/4 & 3.8/4 & -- \\
 stat.\ err.\ & 0.0002 & 0.0055 & 0.0108 & 0.0144 & 0.0065 & 0.0032 & 0.0034 & 0.0034 & 0.0015 \\
 syst.\ err.\ & 0.0011 & 0.0011 & 0.0011 & 0.0011 & 0.0012 & 0.0013 & 0.0014 & 0.0011 & 0.0013\\
 hadr.\ err.\ & 0.0044 & 0.0028 & 0.0022 & 0.0021 & 0.0019 & 0.0018 & 0.0017 & 0.0016 & 0.0022\\
 pert.\ err.\ & $^{+0.0006}_{-0.0011}$ & $^{+0.0006}_{-0.0011}$ & $^{+0.0009}_{-0.0014}$ &   $^{+0.0003}_{-0.0005}$ &
                $^{+0.0007}_{-0.0011}$ & $^{+0.0006}_{-0.0009}$ & $^{+0.0006}_{-0.0008}$ &   $^{+0.0005}_{-0.0007}$ & 0.0009 \\
 soft.\ err.\ & 0.0005 & 0.0005 & 0.0006 & 0.0002 & 0.0005 & 0.0004 & 0.0004 & 0.0004 & 0.0004\\
tot.\ err.\  & 0.0047 & 0.0064 & 0.0112 & 0.0147 & 0.0070 & 0.0040 & 0.0041 & 0.0040 & 0.0031 \\
\hline
 {$\alpha_s(m_Z)$} & 0.1214 & 0.1235 & 0.1328 & 0.1077 & 0.1267 & 0.1234 & 0.1218 & 0.1189 &0.1220 \\
\hline
\hline
{\sc pythia}  & 0.1365 & 0.1239 & 0.1333 & 0.1073 & 0.1266 & 0.1214 & 0.1202 & 0.1168 & 0.1230 \\
{\sc ariadne} & 0.1238 & 0.1262 & 0.1355 & 0.1093 & 0.1288 & 0.1239 & 0.1731 & 0.1687 & 0.1250 \\
\hline
  \end{tabular}
  \caption{Best fit to {\sc aleph} data. The row labelled ``pert err.'' is derived from scale uncertainties
and the row labelled ``soft err.'' from the uncertainty on $\ctwo$ and $\ctwoL$ in Eq.\eqref{ourchoice}.
\label{tab:alephresults}
The rows labeled {\sc pythia}  and {\sc ariadne}  give 
the value of $\alpha_s$ after correcting for hadronization
and quark masses using {\sc pythia} or {\sc ariadne}. The {\sc ariadne} corrected prediction
for the two highest two energies produce very poor fits, and are excluded from the average}
\end{center}
\end{table}

\begin{figure}[t]
\begin{center}
\includegraphics[width=0.7\textwidth]{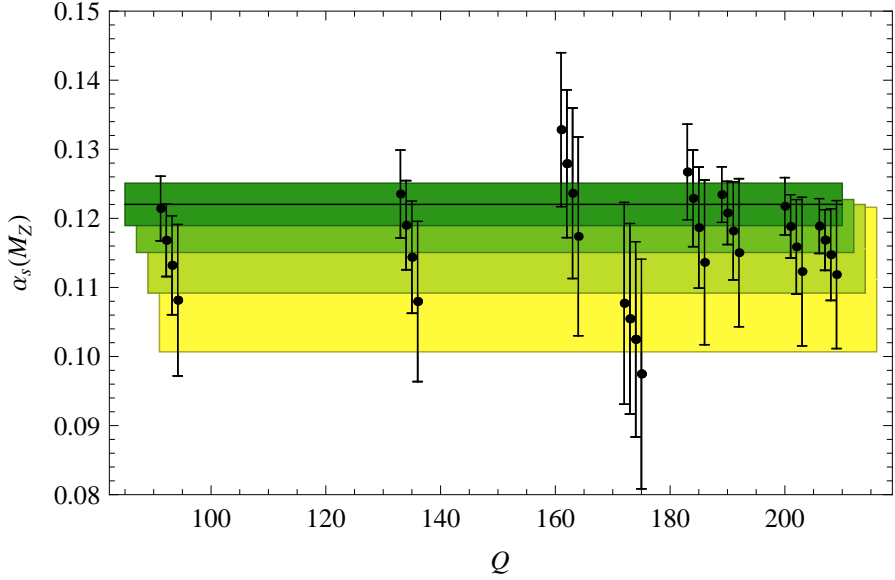}
\end{center}
\vspace{-0.5cm}
\caption{Best fit values for $\alpha_s(m_Z)$. From right to left the lines are the total error bars at each energy for
 first order, second order, third order and fourth order, as defined in the text. The bands are weighted averages
with errors combined from all energies.}
\label{fig:fitresults}
\end{figure}
We show in Figure~\ref{fig:fitresults} the convergence of the best fit values as a function of energy.
There is very good consistency among the different energies and the convergence order-by-order in perturbation theory
is good as well. The fit values for different orders are given in Table~\ref{tab:fitorder}.

\begin{table}
\begin{center}
 \begin{tabular}{|c|c|c|c||c|c|c|}  \hline
 &  \multicolumn{3}{|c||}{\lepone +\leptwo} & \multicolumn{3}{c|}{\lepone ($91.2$ GeV only)}  \\ \hline
    order  & $\alpha_s$ & total err  & pert. err & $\alpha_s$  & tot.err & pert.err  \\  \hline
    \first order & 0.1111 & 0.0104 & 0.0100 & 0 1099 & 0.0100 & 0.0110 \\
    \second order& 0.1156 & 0.0064 & 0.0057 & 0.1132 & 0.0072 & 0.0055 \\
    \third order & 0.1189 & 0.0038 & 0.0025 & 0.1168 & 0.0052 & 0.0026 \\
    \fourth order& 0.1220 & 0.0031 & 0.0009 & 0.1214 & 0.0047 & 0.0011 \\ \hline
  \end{tabular}
  \caption{Best fit values and uncertainties at different orders.}
  \label{tab:fitorder}
\end{center}
\vspace{-0.5cm}
\end{table}

The final fit for heavy jet mass gives
 \begin{align}
  \alpha_s (m_Z) &= 0.1220 \pm 0.0014~\text{(stat)} \pm 0.0013~\text{(syst)} \pm
  0.0022~\text{(had)} \pm 0.0009~\text{(pert)} \pm 0.0004~\text{(soft)} \nonumber \\
& = 0.1220 \pm 0.0031\, \quad  \text{\bf (Heavy Jet Mass)}\,.
\end{align}
This can be compared to the result for thrust, using exactly the same technique,
and the same energy {\sc aleph} data (Table 2 of~\cite{Becher:2008cf}). Updating this result
to include the more recent NNLO distributions~\cite{GehrmannDeRidder:2007hr,Weinzierl:2008iv},
 using the same $\ctwo$ values, 
Eq.\eqref{ourchoice}, with associated ``soft'' uncertainty, and restricting to only the {\sc aleph} data, we find
 \begin{align}
  \alpha_s (m_Z) &= 0.1175 \pm 0.0009~ (\mathrm{stat}) \pm 0.0011~ (\mathrm{syst}) \pm
  0.0014~ (\mathrm{had}) \pm 0.0016~ (\mathrm{pert}) \pm 0.0006~  (\mathrm{soft}) \nonumber \\
& = 0.1175 \pm 0.0026\, \quad \text{\bf  (Thrust)}\,.
\end{align}
Combining these results, assuming $100\%$ correlation between heavy jet mass and thrust, gives
 \begin{align}
  \alpha_s (m_Z) &= 0.1193 \pm 0.0011~ (\mathrm{stat}) \pm 0.0012~ (\mathrm{syst}) \pm
  0.0017~ (\mathrm{had}) \pm 0.0013~ (\mathrm{pert}) \pm 0.0005~  (\mathrm{soft}) \nonumber \\
& = 0.1193 \pm 0.0027\, \quad  \text{\bf (Combined)}\,.
\end{align}
This value is consistent with the recent world average of $\alpha_s(m_Z) = 0.1184 \pm 0.0007$~\cite{Bethke:2009jm}.


\section{Non-perturbative effects and quark mass corrections \label{sec:NP}}
The  $\alpha_s$ fit from the previous section used the theory prediction at the parton level
with five flavors of massless quarks, neglecting hadronization and quark masses. 
Hadronization induces a power correction on the heavy jet mass distribution. Its effect is suppressed by a small scale,
such as  $\LQCD/Q$ or $\LQCD/{\mu_s}$ relative to the perturbatively calculable part. 
The $b$-quark mass corrections are suppressed by $m_b/Q$.
These effects are therefore both
parametrically smaller than the large logarithmic corrections which we resum. Nevertheless, they
are quantitatively important, and our final uncertainty on $\alpha_s$ is dominated
by the way these power corrections are modeled. 
The dominant part of the $b$-quark mass corrections is calculable,
and is expected to shift $\alpha_s$ at around the 1\% level, as observed in~\cite{Becher:2008cf,Abbate:2010vw}. 
The inclusion of $b$-quark corrections will be an important addition for future work. However, since they scale like $1/Q$,
the dominant effect of these mass corrections can be absorbed into the same power correction model as hadronization effects,
which also scale as $1/Q$. 
In this section, we
explore the Monte Carlo treatment of power corrections, and an alternative theoretical model.

Monte Carlo simulations can include quark masses explicitly. They also attempt to model hadronization, for example
with a string fragmentation model in {\sc pythia}. This produces an event with stable particles
which can be run through a detector simulation. Such simulations are an essential part of every experimental
study, and must play some role even for inclusive event shape analysis. For example, the event shape is
often measured using only the charged particles, whose momenta are more precisely known,
and then corrected to all particles with help of the simulation.
Monte Carlo hadronization models have a number of free parameters
and can usually be tuned to any particular data set so that the simulation reproduces the data quite well. 
However, no single tuning reproduces all the data, and therefore different tunings are often used for different
analyses. A more troubling fact is that, as demonstrated in~\cite{Becher:2008cf},
the tunings often correct for features having nothing to do with hadronization, such as subleading log resummation.
Such tunings are guaranteed not to scale well with energy. This may be a serious problem for high energy colliders
which simultaneously probe many energy scales, such as the Large Hadron Collider at CERN.

The hadronization  uncertainty we used for the $\alpha_s$ determination
were taken from~\cite{Dissertori:2007xa}, but we have also studied hadronization and mass effects in the Monte Carlos on our
own.  The last two rows of Table~\ref{tab:alephresults} show the best fit vales for $\alpha_s$ after the theory is
corrected bin-by-bin for both hadronization and bottom and charm quark mass corrections
using the Monte Carlo event generators {\sc pythia v6.409}, with
default parameters~\cite{Sjostrand:2006za}, and {\sc ariadne v4.12} with the {\sc aleph} tune~\cite{Lonnblad:1992tz}. Recall
that {\sc ariadne} actually feeds through {\sc pythia} to handle hadronization, so the difference is entirely due to the way
the parton shower is implemented. With thrust, the same exercise was performed, and
the corrections with {\sc ariadne} were found to be very small, which helped
justify not correcting for hadronization and quark masses at all in the published $\alpha_s$ fits. For thrust corrected with {\sc pythia},
there was a systematic downward shift in $\alpha_s$.
For heavy jet mass, the corrections with {\sc ariadne} are large. In fact, for the high energy data, ridiculous values 
such as $\alpha_s=0.1731$ result. The {\sc pythia} corrections are, for heavy jet mass, smaller than they are for
thrust. In fact, we find a bigger discrepancy between
the thrust and heavy jet mass $\alpha_s$ fits after correcting with either Monte Carlo than without.
Thus, although we cannot justify correcting the theory curve with either Monte Carlo,
we confirm that the hadronization uncertainties listed in Table~\ref{tab:alephresults}, which were taken
from~\cite{Dissertori:2007xa}, span reasonable Monte-Carlo simulated variations due to hadronization
and quark mass effects.

\begin{figure}[t]
\psfrag{Ry}[l]{$\frac{\rd \sigma}{\rd \rho}$}
\psfrag{Rx}[]{$\rho$}
\psfrag{Tw}[l]{$\frac{\rd \sigma}{\rd \tau}$}
\psfrag{Tz}[]{$\tau$}
\begin{center}
\includegraphics[width=0.9\textwidth]{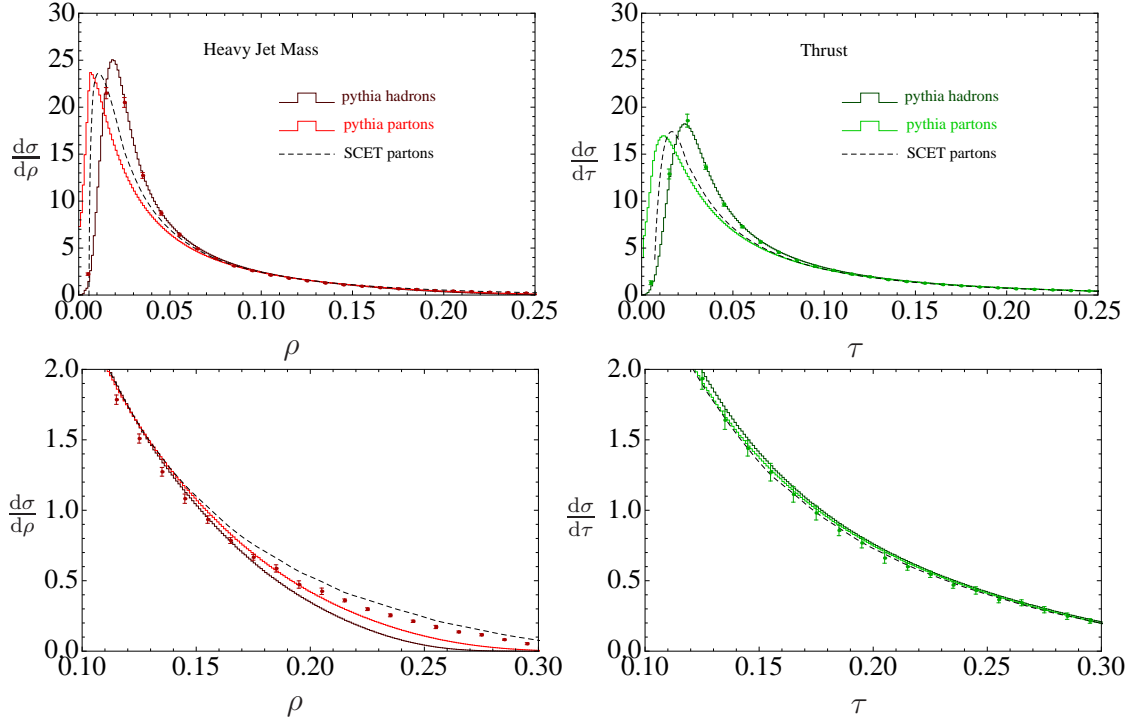}
\end{center}
\caption{Hadronization and mass corrections with {\sc pythia}. The theoretical prediction using {\sc pythia}
at the hadron level with massive quarks and the parton level with massless quarks
 is compared to data and to the \fourth order theoretical prediction using SCET.
For thrust, {\sc pythia} agrees remarkably well with the data, while for heavy jet mass, there is a substantial discrepancy
especially in the fit region, which is zoomed in on in the bottom panels.
 \label{fig:pythia}}
\end{figure}
To understand why the power corrections come out so differently for thrust and heavy jet mass, we compare {\sc pythia}
at the parton and hadron levels to the \fourth order SCET prediction (N${}^3$LL + NNLO), and to the {\sc aleph} data at 91.2 GeV in Figure~\ref{fig:pythia}.
From the top two panels, we see that in the peak region, in both cases the parton-level theory prediction comes out somewhere
between the parton and hadron level Monte Carlo. However, in the bottom two panels, which zoom in near the fit region, the difference
between the two event shapes is much more dramatic. For heavy jet mass, the SCET curve is above the data, while partonic {\sc pythia}
is below it and hadronic {\sc pythia} is even farther below. In contrast, for thrust, all of the curves
are much closer and the power corrections, as modeled by {\sc pythia} are a much smaller effect. 
It is clear that {\sc pythia}
has trouble handling both event shapes simultaneously.

An alternative to using Monte Carlo simulations to simulate hadronization is to model the power corrections directly with 
effective field theory. As discussed in~\cite{Hoang:2008fs}, hadronization effects can be absorbed into the soft function
by convolution of the perturbatively calculable part with a non-perturbative shape function
\begin{equation}
  S_{\mathrm{full}}(k_L, k_R,\mu) = \int \rd k_L' \rd k_R' S_{\mathrm{part}}(k_L -k_L',k_R - k_R',\mu) S_{\mathrm{mod}}(k_L',k_R') \,,
\end{equation}
where $S_{\mathrm{part}}(k_L,k_R,\mu)$ is what we have previously just been calling $S(k_L,k_R,\mu)$ and $S_{\mathrm{mod}}(k_L,k_R)$
is a non-perturbative model function. Generally,  $S_{\mathrm{mod}}(k_L,k_R)$ is expected to have support only for 
$k_L,k_R \lesssim \Lambda_{\mathrm{QCD}}$. As observed in~\cite{Hoang:2007vb}, 
there is an ambiguity in the factorization of the soft function into
these two pieces, which leads to a difficulty in assigning physical significance to $S_{\mathrm{mod}}(k_L,k_R)$
and poor convergence in perturbation theory. This ambiguity is associated with the existence of a renormalon,
which can be removed within SCET~\cite{Hoang:2007vb,Hoang:2008fs}.  Indeed, if data closer to the peak region were included in the fit, 
or if convergence of the model function parameters were an issue, removing the renormalon could have an important effect.
Since we are not immediately interested in these issues, for simplicity, 
we will simply ignore the renormalon.

The simplest model function is just composed of delta functions
\begin{equation}
  S_{\mathrm{mod}}(k_L,k_R) = \delta(k_L -\frac{1}{2} \Lnp) \delta(k_R -\frac{1}{2} \Lnp) \,.
\end{equation}
The one parameter, $\Lnp$, can be thought of as representing the mass gap of QCD due to hadronization
and therefore should be of order $\LQCD$. This model function allows us to fit the leading power correction.
 Any other one-parameter family of shape functions
can be written in this form up to higher power corrections, which should have a subleading effect on the distributions.
For example, the smallest scale probed in our fits is the soft scale at the lower end of the fit region at 91.2 GeV, 
 $\mu = \mu_s= \rho Q >$ (0.08)(91.2 GeV) $\sim 7$ GeV.
With $\LQCD\sim 300$ MeV, this can be a 4\% effect. Higher power corrections, of order $(\LQCD/\mu)^2$ should have
less than a 0.2\% effect in our fit range.

\begin{figure}[t]
\psfrag{x}[]{\small $\alpha_s(m_Z)$}
\psfrag{y}[]{\small $\LNP$ (GeV)}
\begin{center}
\includegraphics[width=0.7\textwidth]{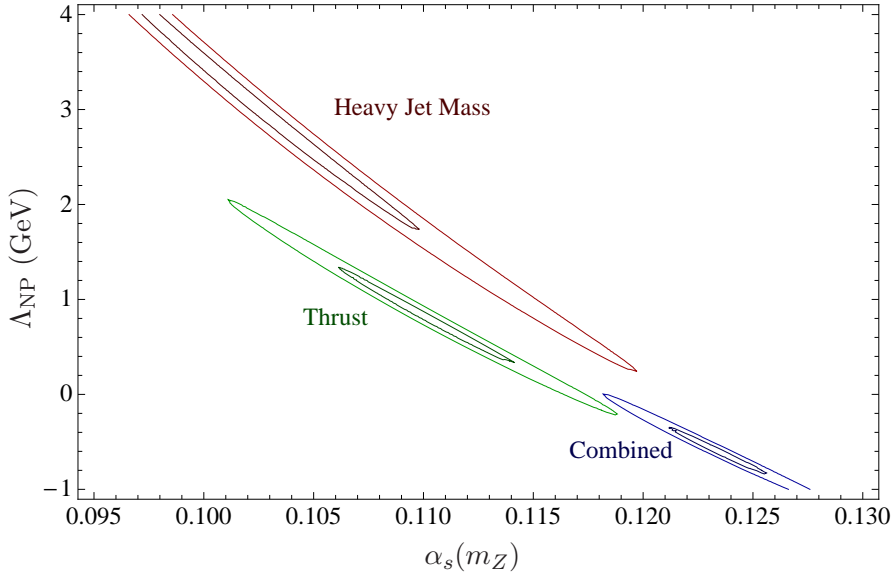}
\end{center}
\caption{Contours of $2 \sigma$ and $5\sigma$ confidence in the simultaneous fit of $\alpha_s$ and a non-perturbative
shift parameter $\Lambda_{\mathrm{NP}}$ to the thrust and heavy jet mass {\sc aleph} data from $91.2$ to $206$ GeV.
The combined fit is also shown.}
\label{fig:NPconts}
\vspace*{0.2cm}
\end{figure}

Once this shape function is convoluted with the perturbative distribution, it has the effect of simply shifting the distributions
\begin{align}
  \frac{\rd \sigma}{\rd \tau} (\tau) &\to 
  \frac{\rd \sigma}{\rd \tau} (\tau -\Lnp) \\
  \frac{\rd \sigma}{\rd \rho} (\rho) &\to 
  \frac{\rd \sigma}{\rd \rho} (\rho -\frac{1}{2}\Lnp) \,.
\end{align}
The factor of $\frac{1}{2}$ is easy to understand. The shift causes each hemisphere mass to increase by $\frac{1}{2}\Lnp$.
Since thrust sums both hemisphere masses, while heavy jet mass measures only one, heavy jet mass feels only half
of the increase.

\begin{table}[t]
\begin{center}
 \begin{tabular}{|c|c|c|c|} 
\hline
  Event Shape  & $\alpha_s(m_Z)$ & $\LNP$ (GeV) & $\chi^2$/d.o.f. \\
\hline
Thrust & 0.1101 & 0.821 & 66.9/47 \\
Heavy Jet Mass & 0.1017 & 3.17 & 60.4/43 \\
Combined & 0.1236 & -0.621 &453/92\\
\hline
  \end{tabular}
  \caption{Best fit values including leading power correction.
The $\chi^2$ is calculated using both statistical and experimental systematic uncertainties.}
  \label{tab:npfit}
\vspace*{-0.5cm}
\end{center}
\end{table}

This model was studied for thrust in~\cite{Becher:2008cf}, where it was found that a larger $\LNP$ can be compensated
for by smaller $\alpha_s$ leading to a flat direction in the two parameter fits. 
We reproduce this result in Figure~\ref{fig:NPconts}.
This figure shows the $2\sigma$ and
$5\sigma$ confidence regions in a combined fit to all of the {\sc aleph} data for thrust from LEP.
 On the same plot, using the same model function, we show the contours for heavy jet mass.
First of all, we observe that the flat direction exists
in both of the data sets. We might have hoped that having two event shapes would remove the ambiguity, but this does not happen.
Second, we see that while the perturbative fit has $\alpha_s$ lower for thrust than for heavy jet mass, with the power corrections,
the value of $\alpha_s$ is higher for thrust, as found in previous studies~\cite{Dissertori:2009ik, Dissertori:2007xa}.
However, when we perform a simultaneous fit to all of the thrust and heavy jet mass degrees of freedom, we get a value
for $\alpha_s$ that is larger than each one separately.
The best fit for thrust, heavy jet mass, and the combined fit are shown in Table~\ref{tab:npfit}.
The fact that the thrust and heavy jet mass contours do not overlap indicate that
a better handling of non-perturbative effects is required.

We conclude that neither correcting the theory curves with a Monte Carlo simulation  nor using a minimal shape function
approach for the leading power correction is satisfactory. The shape function approach is improvable, while the Monte Carlo approach
is limited by the perturbative accuracy of the parton shower, which will be limited to leading-log resummation in at least the near future
(although SCET may eventually help go beyond LL~\cite{Bauer:2006mk,Bauer:2006qp}).
To improve the shape function fit, a number of additional ingredients should be included. First of all, the renormalon ambiguity in
separating the perturbative and non-perturbative parts of the soft function should be removed. 
This is not likely to have much effect in the fit
region we use, but will allow us to use data closer to the peak. Having more data involved will more highly constrain the fit and could 
remove the flat direction. To do this, we would need the bin-by-bin correlations among the experimental systematic uncertainties,
which are not publicly available.
In addition, there are perturbatively calculable effects we have not included,  
such as electroweak and $m_b$ corrections, as in~\cite{Abbate:2010vw}, which may have up to a 1\% effect.
It would be very interesting to see if the thrust and heavy jet mass distributions can
be reconciled once a thorough effective field theory analysis, including non-perturbative effects, is performed.

\section{Conclusions \label{sec:conc}}
In this paper, We have studied the heavy jet mass distribution using Soft-Collinear Effective Theory including N${}^3$LL resummation
and matching to the NNLO fixed order distribution. Up to this point, this kind of accuracy has only been available
for the thrust distribution. Having an additional event shape helps control for systematic uncertainties, making the fit
for $\alpha_s$ more trustworthy. It also gives us insight into power corrections and multi-scale soft functions which 
will be important for the LHC.

The heavy jet mass fit gives $\alpha_s(m_Z) = 0.1220 \pm 0.0031$. This value is larger than what had been found for thrust
at the same accuracy, $\alpha_s(m_Z) = 0.1175 \pm 0.0026$.
The uncertainty on heavy jet
mass is larger partially due to a larger hadronization uncertainty.
In our study, no corrections were made for hadronization. 
We explored the traditional method of hadronization, using Monte Carlo event generator, such as {\sc pythia}
and concluded that such an approach is problematic 
for theoretical calculations of this accuracy. Since the Monte Carlo has been already tuned
to the data we are trying to match, the tuning has partially compensated for resummation of subleading logarithms.
Comparing {\sc pythia}'s output in the fit region, the hadronized
distribution is actually farther away from the data than the parton-level distribution.

Our $\alpha_s$ values from thrust and heavy jet mass contrast with the results of~\cite{Dissertori:2009ik},
 which at NLL+NNLO accuracy derived $\alpha_s(m_Z) = 0.1266$ from
thrust and $\alpha_s(m_Z)=0.1211$ from heavy jet mass. A comparison of various fits to thrust and heavy jet mass
using the same {\sc aleph} data is shown in Table~\ref{tab:fits}.
The authors of~\cite{Dissertori:2009ik} have observed that event shapes tend to belong to one of two classes. The first class,
including thrust, tends to produce higher values of $\alpha_s$ than the second class, which includes heavy jet mass.
These authors attributed the difference to better perturbative stability in the second class. 
We find, if anything, better perturbative stability for thrust. Instead,
the reason for the systematic separation of $\alpha_s$ values in this study,
and also in the NNLO study of~\cite{Dissertori:2007xa}, may have more to do with their use
of a Monte Carlo simulation to correct for hadronization. A similar conclusion was reached in~\cite{Gehrmann:2009eh}
which studied event shape moments.
The values of $\alpha_s$ for the two classes must eventually be reconcilable, but there may be a physical
reason why the power corrections for one class are larger than for the other. 
This is worth understanding more thoroughly, and may have implications for the LHC.

\begin{table}[t]
\begin{center}
 \begin{tabular}{|c|c|c|c||c|c|} 
\hline
Order & N${}^3$LL+NNLO & N${}^3$LL+NLO &  NNLO    & NNLO~\cite{Dissertori:2009ik} &NLL+NNLO~\cite{Dissertori:2009ik} \\
\hline
hadronization & NO & NO & NO & YES & YES \\
\hline
      Thrust         & 0.1175      & 0.1173  &  0.1262       & 0.1275 & 0.1266 \\
      Heavy Jet Mass & 0.1220      & 0.1189  &  0.1265       & 0.1248  & 0.1211\\
\hline
  \end{tabular}
  \caption{Best fit values for $\alpha_s(m_Z)$ at various orders in perturbation theory. 
The first three columns are our results,
the last two which include a Monte Carlo based hadronization correction are from~\cite{Dissertori:2009ik}.}
  \label{tab:fits}
\end{center}
\vspace*{-0.5cm}
\end{table}

The alternative to using a Monte Carlo simulation for hadronization is to add a shape function contribution
within the effective field theory.
Our simple shape function study shows that the leading power tends to shift $\alpha_s$ from
both heavy jet mass and from thrust to lower values, with the heavy jet mass shift of larger magnitude.
This can help explain why the thrust $\alpha_s$ comes out lower than the heavy jet mass $\alpha_s$ in our study,
and not in~\cite{Dissertori:2009ik, Dissertori:2007xa}. However, we also found that the best fit over all
the {\sc aleph} data from 91.2 to 206 GeV for thrust was incompatible with the best fit from heavy jet mass,
and that the flat direction between $\alpha_s$ and the non-perturbative parameter $\LNP$ persists in both
distributions. 

To get the values of $\alpha_s$ extracted from thrust and heavy jet mass to agree
may require including additional ingredients, which can be done within 
the effective field theory framework. For example, there
is a calculable $m_b$ correction which tends to bring $\alpha_s$ up at least for thrust~\cite{Abbate:2010vw}.
Including every possible correction  must produce the same value of $\alpha_s$ from thrust and heavy jet mass, and it will
be interesting to see precisely how this happens.
Also, more data should be included. Using data for values of heavy jet mass and thrust closer to the
peak will lead to a more constrained shape function fit, although it may require going beyond the leading power.
In addition, using data from other {\sc lep} experiments and other experiments
at lower center-of-mass energy can further test and constrain the event shapes.

However, it is not clear if all of the differences between thrust and heavy jet mass can be accounted for entirely
within SCET. For example, there is the possibility that the difference between thrust and heavy jet mass has more
to do with the way hadron masses are handled experimentally than from higher order power corrections. In~\cite{Salam:2001bd},
substantial differences in the form of power corrections among the $E$-scheme, $p$-scheme and decay-scheme were found.
It may turn out that an ultra-precise $\alpha_s$ fit can only be made if the identity of all the hadrons is known,
which may be possible for future measurements but is not available for existing data. More likely,
the thrust and heavy jet mass distributions can be made to agree within SCET, but the uncertainty on $\alpha_s$ will
ultimately be limited by a hadron-mass-scheme dependent uncertainty. In any case, once the ingredients discussed
for thrust in~\cite{Abbate:2010vw} are applied to heavy jet mass, we will be able to extract a more precise lesson about
the importance of power corrections. 
In addition to reducing the uncertainty from $\alpha_s$ and teaching us about power corrections,
combining the insights from thrust and heavy jet mass will more generally pave the way for deeper understanding of
relevant jet-based observables at the LHC.

\newpage

\section*{Note Added}
\begin{figure}[t]
\begin{center}
\psfrag{x}{$\rho$}
\psfrag{yyyyyyyy}{$\frac{1}{\sigma_0} \Delta\left[\rho \frac{\rd \sigma}{\rd \rho}\right]$}
\psfrag{yyyyyyyz}{}
\psfrag{CF}{$C_F$}
\psfrag{CA}{$C_A$}
\psfrag{NF}{$n_F$}
\psfrag{total}{$\text{total}$}
\psfrag{cut}{$\text{cutoff}=10^{-12}$}
\includegraphics[width=0.8\textwidth]{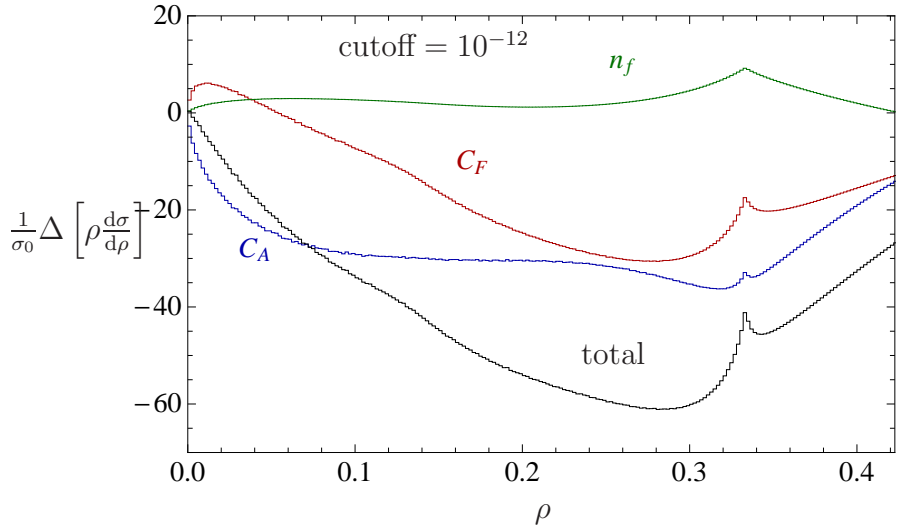}
\end{center}
\vspace*{-0.5cm}
\caption{A comparison of the full fixed-order calculations and expanded SCET at NLO. Update of
Figure~\ref{fig:nlodiff} with cutoff $y=10^{-12}$ in $B(\rho)$ from {\sc event 2}.}
\label{fig:nlodiffcut12}
\end{figure}

\begin{figure}[t]
\begin{center}
\psfrag{y}[]{$\ctwor$}
\psfrag{x}[]{$\rmin$}
\includegraphics[width=0.7\textwidth]{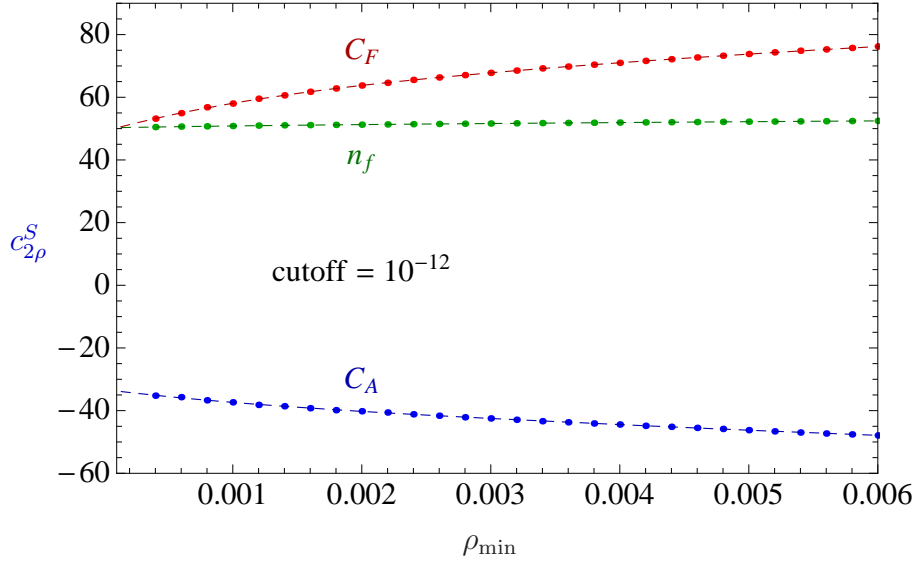}
\end{center}
\vspace*{-0.5cm}
\caption{Extraction of the two-loop constants in the soft function. Update of Figure~\ref{fig:softNLO} with cutoff $y=10^{-12}$
in $B(\rho)$ from {\sc event 2}.}
\label{fig:softNLOcut12}
\end{figure}

After this manuscript appeared, it was suggested that the precision on $\ctwor$ and $\ctwo$ could be improved by lowering
the infrared cutoff used by {\sc event 2}. The cutoff $y$ is implemented by throwing out events if two partons have
$(p_1 + p_2)^2 < y Q^2$. The default cutoff is $10^{-8}$, and the authors of {\sc event 2} caution about numerical instabilities
if the cutoff is taken too low. We find that for cutoffs below $10^{-15}$, there are insurmountable numerical problems, however
$y=10^{-12}$ seems to be convergent. We therefore ran 135 billion events with $y=10^{-12}$ and 2500 bins ($\Delta \rho = 0.0002)$)
-- the main text uses 10 billion events with $y=10^{-8}$ and 1000 bins ($\Delta \rho = 0.0005$). The difference between
this new numerical data and the SCET prediction for the singular terms is shown in Figure~\ref{fig:nlodiffcut12}, which is to
be compared to Figure~\ref{fig:nlodiff}. One can see that the curves for all color structures now converge to zero, as expected.

Next, the constant $\ctwor$ was extracted from these curves. The value $\ctwor$ for various lower bounds $\rmin$ are
shown in Figure~\ref{fig:softNLOcut12}. Again, improved numerical stability is clear. 
Fitting a sixth order polynomial to the 59 points between $\rmin=0.0004$ and $\rmin = 0.012$ and extrapolating to $\rmin=0$
leads to
\begin{equation}
   \ctwor =(49.1) C_F^2 + (-33.2) C_F C_A + (50.2)C_F T_F n_f \,. \label{ctwornew}
\end{equation}
The value of the $C_F^2$ coefficient is now consistent with the prediction of $\frac{\pi^4}{2} = 48.7$ from non-Abelian exponentiation.
The fit is somewhat sensitive to the lower value of $\rmin$ used in the regression, but not very sensitive to the upper value.
Fitting a fourth order polynomial to the 38 points between $\rmin=0.0006$ and $\rmin=0.008$ gives
$\ctwor =(49.8) C_F^2 + (-33.3) C_F C_A + (50.3)C_F T_F n_f$. Since the $C_F C_A$ and $C_F T_F n_f$ terms are practically unchanged,
and the $C_F^2$ term is fixed by non-Abelian exponentiation, it is reasonable to assume that the remaining uncertainty on
these numbers will have a negligible effect on the $\alpha_s$ fits. 
Performing the same analysis for thrust, leads to
\begin{equation}
   \ctwo =(49.1) C_F^2 + (-57.8) C_F C_A + (43.4)C_F T_F n_f
\end{equation}
Combining these, assuming the Hoang-Kluth Ansatz for the soft function, Eq.~\eqref{softansatz}, gives
\begin{equation}
   \ctwoL =(0) C_F^2 + (-7.5) C_F C_A + (-2.1)C_F T_F n_f 
\end{equation}
Thus, we take
\begin{align}
   \ctwo  &=\frac{\pi^4}{2} C_F^2 + (-57.8) C_F C_A + (43.4)C_F T_F n_f \label{ctwoup}\\
   \ctwoL &= (0) C_F^2 + (-7.5) C_F C_A + (-2.1)C_F T_F n_f  
\end{align}

\begin{figure}[h!]
\begin{center}
\psfrag{N2}[l]{\small $N^2$}
\psfrag{N0}[l]{\small $N^0$}
\psfrag{Nm}[l]{\small $1/N^{2}$}
\psfrag{Nnf}[l]{\small $n_f N$}
\psfrag{NfNm}[l]{\small $\phantom{a}n_f/N$}
\psfrag{Nf2}[l]{\small $n_f^{2}$}
\psfrag{sig}[t]{ $\frac{10^{-3}}{\sigma_0} \rho \frac{\rd\sigma}{\rd \rho}$}
\psfrag{x}[b]{\small $\phantom{abc}-\log\rho$}
\psfrag{cutoff}[t]{$\text{cutoff} = 10^{-7}$}
\includegraphics[width=0.85\textwidth]{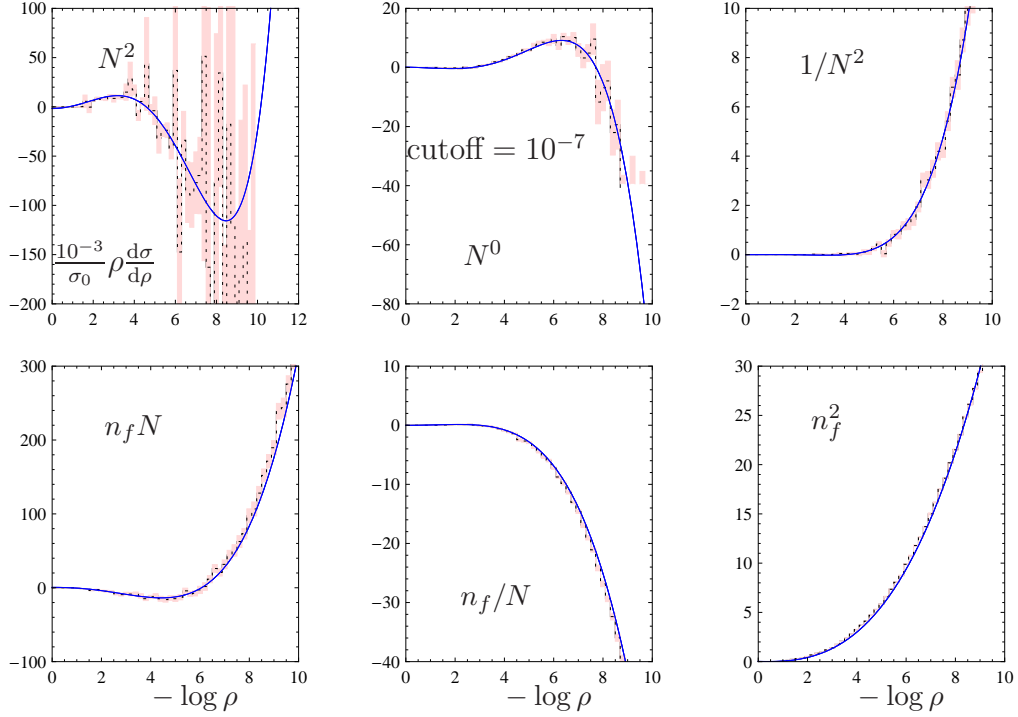}

\end{center}
\vspace*{-0.9cm}
\caption{\label{fig:NNLOcolslogcut7}
Comparison of the full NNLO heavy jet mass distribution 
and the singular terms. Update of Fig~\ref{fig:NNLOcolslog} with cutoff $y_0 = 10^{-7}$ in $C(\rho)$.
}
\end{figure}

\begin{figure}[h!]
\begin{center}
\vspace{-0.1cm}
\psfrag{N2}[l]{\small $N^2$}
\psfrag{N0}[l]{\small $N^0$}
\psfrag{Nm}[l]{\small $1/N^{2}$}
\psfrag{Nnf}[l]{\small $n_f N$}
\psfrag{NfNm}[l]{\small $n_f/N$}
\psfrag{Nf2}[l]{\small $n_f^{2}$}
\psfrag{sig}[t]{$\frac{1}{\sigma_0} \Delta\left[\rho \frac{\rd \sigma}{\rd \rho}\right]$}
\psfrag{cutoff}[t]{$\text{cutoff} = 10^{-7}$}
\psfrag{x}[b]{\small $\rho$}
\includegraphics[width=0.85\textwidth]{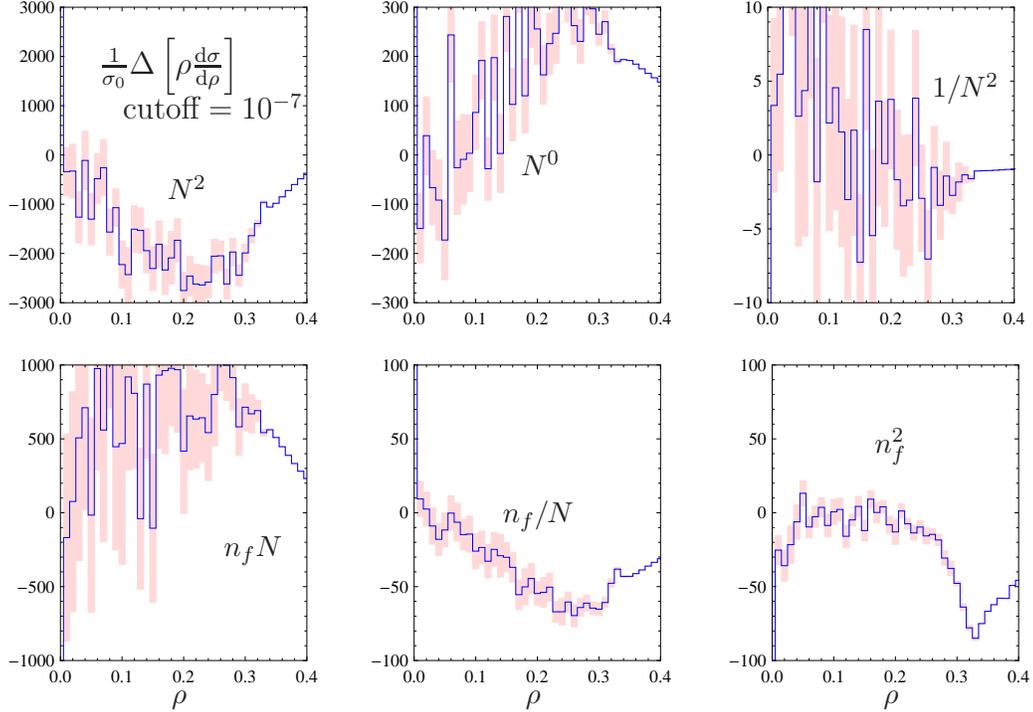}
\end{center}
\vspace*{-0.9cm}
\caption{\label{fig:nnlodiffcut7}  
Difference between full NNLO heavy jet mass distribution and the singular terms.
Update of Fig~\ref{fig:nnlodiff} with cutoff $y_0 = 10^{-7}$ in $C(\rho)$.
The uncertainty on $\ctwo$ and $\ctwor$ is now negligible.
}
\end{figure}

With these more accurate numbers and a more accurate numerical calculation of the NNLO distribution, we can
now repeat our comparison of the singular terms to the exact distribution. Using an infrared cutoff of $10^{-7}$ for
the $C$ functions, the agreement with the singular terms is improved. This can be seen in Figure~\ref{fig:NNLOcolslogcut7},
which is an update of Figure~\ref{fig:NNLOcolslog}. Taking the difference between the curves gives Figure~\ref{fig:nnlodiffcut7}.
One sees that the $1/N^2$ color structure, corresponding to $C_F^3$, has improved convergence towards zero. If these
curves were known with perfect accuracy, they could be used to test the Ansatz in Eq.~\eqref{softansatz}. 
The most poorly convergent color structures, $1/N^2$ and $n_f^2$ are not sensitive to this Ansatz, and the others 
are consistent with convergence to zero within the statistical uncertainty on the numerical NNLO calculation.

Finally, we reconsider the $\alpha_s$ fits in light of these more precise soft function coefficients and NLO matching
functions. Refitting the
thrust distribution to the {\sc aleph} data
changes $\alpha_s(m_Z)$ from $0.1175$ to $0.1176$ and refitting the heavy jet mass distribution raises $\alpha_s(m_Z)$
from $0.1220$ to $0.1224$. These shifts are within the quoted soft uncertainties.


\subsection*{Acknowledgements}
The authors would like to thank T. Becher and T. Gehrmann for discussions and comments on this manuscript. We also
thank T. Gehrmann for providing us with the NNLO distributions, and for helping us understand the discrepancies
in Figures~\ref{fig:NNLOcolslog} and~\ref{fig:nnlodiff}.
We would also like to thank Andre Hoang and Gavin Salam for helpful discussions.
Our research is supported in part by the Department of Energy OJI program, under Grant DE-AC02-76CH03000.


\begin{appendix}
\section{Soft function \label{app:soft}}
To evaluate the heavy jet mass distribution with NNLO precision, we need the soft function at the scale $\mu_s$ evaluated to
order $\alpha_s^3$.  The Laplace transformed soft function can be written as
\begin{equation}
  \widetilde{s}({\rLO},{\rLT}) =
\widetilde{s}_\mu({\rLO},\mu)
\widetilde{s}_\mu({\rLT},\mu) 
 \tf( {\rLO} - {\rLT} ) \,,
\end{equation}
where $\rLO = \ln(\mu {\red  \nu_L} e^{\gamma_E})$ and $\rLT = \ln(\mu {\red \nu_R} e^{\gamma_E})$,
with ${\red \nu_L}$ and ${\red \nu_R}$ the Laplace conjugate variables to the soft momenta $k_L$ and $k_R$. 
The $\mu$-dependence is determined by the function we call
$\wt{s}_\mu(\rL,\mu)$. This is equivalent to the function $U_s(x,\mu,(i x e^{\gamma_E})^{-1})$ in~\cite{Hoang:2008fs},
and we have already given its expansion to order $\alpha_s^3$ in Eq.~\eqref{smuexp}.
The function $\tf(\rL)$ is $\mu$-independent with  $\alpha_s$ evaluated 
at the scale $({\red \nu_L}{\red \nu_R})^{-1/2}$.
It is more useful to be able to use $\alpha_s$ evaluated at the scale $\mu$, which we can do with the replacement
\begin{equation}
\left(\frac{\alpha_s}{4\pi}\right) \to
  \left(\frac{\alpha_s(\mu)}{4\pi}\right)+ \left(\frac{\alpha_s(\mu)}{4\pi}\right)^2
\left[ -\beta_0(\rLO+ \rLT)\right]
+\left(\frac{\alpha_s(\mu)}{4\pi}\right)^3
\left[ \beta_0^2 (\rLO +\rLT)^2-\beta_1(\rLO +\rLT)\right] \,.
\end{equation}
Then truncating the soft function to finite order will induce some residual $\mu$-dependence.

For the finite part $\tf(\rL)$, we use for numerical studies the form in Eq.~\eqref{softansatz}:
\begin{equation} \label{softa2}
  \tf(\rLM) =  1 + \left(\frac{\alpha_s}{4\pi}\right) \cone + \left(\frac{\alpha_s}{4\pi}\right)^2 
\left[\ctwo +\ctwoL\rLM^2\right]+\cdots \,,
\end{equation}
with
\begin{equation}
  \cone = -C_F\pi^2
\end{equation}
and 
\begin{align}
   \ctwo  &=\frac{\pi^4}{2} C_F^2 + (-57.8) C_F C_A + (43.4)C_F T_F n_f \\
   \ctwoL &= (0) C_F^2 + (-7.5) C_F C_A + (-2.1)C_F T_F n_f  
\end{align}
which have been extracted using SCET from the exact NLO thrust and heavy jet mass 
distributions. 
(Note: these numbers are updated to Eq.~\eqref{ctwoup}. The main text and fits use the earlier values in Eq.~\eqref{ourchoice}.)

More generally, for NLO-matching, all that is relevant is a single projection of the order $\alpha^2$ soft function
\begin{equation}
  \ctwor  =\frac{1}{\pi} \int_0^\pi \tf{}_2(i \rL)\, \rd \rL= \ctwo-\frac{\pi^2}{3} \ctwoL  \,.
\end{equation}
For NNLO matching, another projection is necessary, of the form in Eq.~\eqref{ctwozdef}. For the form in Eq.~\eqref{softa2}, this 
projection is
\begin{equation}
  \ctwoz  =\frac{2}{\pi} \int_0^\pi \tf{}_2(i \rL)\ln\left[2\cos(\frac{\rL}{2})\right]\rd \rL =4 \zeta_3 \ctwoL \,.
\end{equation}
These are then expanded as
\begin{equation}
  \ctwo = C_F^2 \ctwocf + C_F C_A \ctwoca + C_F n_F T_F \ctwonf \,.
\end{equation}
We will use these expressions for the singular heavy jet mass expansion and the $G_{ij}$ coefficients below.
 We also use anomalous dimensions and $\beta$-function coefficients 
which can be found in~\cite{Becher:2008cf}.

\section{Expanded soft function}
Putting the pieces together, the soft function expanded to order $\alpha_s^3$ with $\alpha_s = \alpha_s(\mu)$ is

\begin{align}
& \widetilde{s}({\rLO},{\rLT},\mu) =
1 + 
  \left(\frac{\alpha_s}{4\pi}\right)
\Big[
    -({\rLO}^2+{\rLT}^2) \Gamma_0 + (\rLO+\rLT) \gamma^S_0 + \cone 
\Big]\nn \\
  &~~~~~+ \left(\frac{\alpha_s}{4\pi}\right)^2
\Big[
  \frac{1}{2}({\rLO}^4 +{\rLT}^4)\Gamma_0^2 + \rLO \rLT ( \rLO \Gamma_0 - \gamma_0^S)(\rLT \Gamma_0 - \gamma_0^S)
  + ({\rLO}^3 + {\rLT}^3)\left(\frac{2}{3}\beta_0\Gamma_0 - \Gamma_0 \gamma_0^S\right)\nn \\
  &~~~~~+\left({\rLO}^2+{\rLT}^2\right)\left(-\Gamma_1 -\beta_0 \gamma_0^S + \frac{1}{2}(\gamma_0^S)^2-\cone \Gamma_0\right)
  +(\rLO +\rLT)(\gamma_1^S-\cone \beta_0 +
 \cone\gamma_0^S)\nn \\
  &~~~~~+\ctwo + (\rLO-\rLT)^2 \ctwoL\Big]\nn\\
&~~~~~+\Big(\frac{\alpha_s}{4\pi}\Big)^3\Big[~-\frac{1}{6}\Gamma_0^3(\rLO^6+\rLT^6+3\rLO^2\rLT^4+3\rLT^2\rLO^4)
  +\Big(-\frac{2}{3}\beta_0\Gamma_0^2+\frac{1}{2}\gamma^S_0\Gamma_0^2\Big)(\rLO^5+\rLT^5)\nn\\
  &~~~~~+\frac{1}{2}\gamma_0^S\Gamma_0^2(\rLO^4\rLT+\rLT^4\rLO)+\Big(-\frac{2}{3}\beta_0\Gamma_0^2+\gamma_0^S\Gamma_0^2\Big)(\rLO^3\rLT^2+\rLT^3\rLO^2)\nn\\
  &~~~~~+\Big(-\frac{1}{2}(\gamma_0^S)^2\Gamma_0+\frac{5}{3}\gamma_0^S\beta_0\Gamma_0-\frac{2}{3}\beta_0^2\Gamma_0+\Gamma_0\Gamma_1\Big)(\rLO^4+\rLT^4)\nn\\
  &~~~~~+\Big(-(\gamma_0^S)^2\Gamma_0+\frac{2}{3}\gamma_0^S\beta_0\Gamma_0\Big)(\rLO^3\rLT+\rLT^3\rLO)+\Big(-(\gamma_0^S)^2 \Gamma_0 +2\gamma_0^S\beta_0\Gamma_0+2\Gamma_0\Gamma_1\Big)\rLO^2\rLT^2\nn\\
  &~~~~~+\Big(\frac{1}{6}(\gamma_0^S)^3-(\gamma_0^S)^2\beta_0+\frac{4}{3}\gamma_0^S\beta_0^2-\gamma_1^S\Gamma_0
  +\frac{2}{3}\beta_1\Gamma_0-\gamma_0^S\Gamma_1+\frac{4}{3}\beta_0\Gamma_1\Big)(\rLO^3+\rLT^3)\nn\\
  &~~~~~+\Big(+\frac{1}{2}(\gamma_0^S)^3-(\gamma_0^S)^2\beta_0-\gamma_1^S\Gamma_0-\gamma_0^S\Gamma_1\Big)(\rLO^2\rLT+\rLT^2\rLO)\nn\\
  &~~~~~+\Big(\gamma_0^S\gamma_1^S-2\gamma_1^S\beta_0-\gamma_0^S\beta_1-\Gamma_2\Big)(\rLO^2+\rLT^2)
  +2\gamma_0^S\gamma_1^S\rLO\rLT+\gamma_2^S(\rLO+\rLT)\nn\\
  &~~~~~+c^S_1\Big\{\frac{1}{2}\Gamma_0^2(\rLO^4+2\rLO^2\rLT^2+\rLT^4)
  +\Big(-\gamma_0^S\Gamma_0+\frac{5}{3}\beta_0\Gamma_0\Big)(\rLO^3+\rLT^3)\nn\\
  &~~~~~+\Big(-\gamma_0^S\Gamma_0+\beta_0\Gamma_0\Big)(\rLO^2\rLT+\rLT^2\rLO)
  +\Big(\frac{1}{2}(\gamma_0^S)^2-2\gamma_0^S\beta_0+2\beta_0^2-\Gamma_1\Big)(\rLO^2+\rLT^2)\nn\\
  &~~~~~+\Big((\gamma_0^S)^2-2\gamma_0^S\beta_0\Big)\rLO\rLT+\Big(-\beta_1+\gamma_1^S\Big)(\rLO+\rLT)\Big\}\nn\\
  &~~~~~+\Big(\ctwo+\ctwoL(\rLO-\rLT)^2\Big)\Big\{(-2\beta_0+\gamma_0^S)(\rLO+\rLT)-\Gamma_0(\rLO^2+\rLT^2)\Big\}\Big]\;.
\end{align}

\section{Singular terms in the heavy jet mass distribution \label{app:DABC}}
The singular part of the heavy jet mass distribution is calculable in SCET. Writing it in the form
\begin{equation}
    D(\rho)=\delta(\rho)D_\delta+\Big(\frac{\alpha_s}{2\pi}\Big)[D_A(\rho)]_+
    +\Big(\frac{\alpha_s}{2\pi}\Big)^2[D_B(\rho)]_++\Big(\frac{\alpha_s}{2\pi}\Big)^3[D_C(\rho)]_++\ldots\;,
\end{equation}
the result is
\begin{align}
&D_\delta
    =1+\left(\frac{\alpha_s}{4\pi}\right)
\left[C_F\left(-2+\frac{2\pi^2}{3}\right)\right]\\
  &~~~~~+\left(\frac{\alpha_s}{4\pi}\right)^2\left[
     C_F^2\left(4+\frac{\pi^4}{10}-48\zeta_3\right)
    +C_AC_F\left(\frac{493}{81}+\frac{85\pi^2}{6}-\frac{73\pi^4}{90}+\frac{566\zeta_3}{9}\right)\right.\\
  &~~~~~+\left.C_FT_Fn_f\left(\frac{28}{81}-\frac{14\pi^2}{3}-\frac{88\zeta_3}{9}\right)
    + \ctwor \right] \, ,
\end{align}
and
\begin{eqnarray}
    D_A(\rho)
    &&=\frac{1}{\rho}\Big\{C_F\Big[-4\ln\rho-3\Big]\Big\}\;,\nn\\
    D_B(\rho)
    &&=\frac{1}{\rho}\Big\{C_F^2\Big[8\ln^3\rho+18\ln^2\rho+(13-\frac{8\pi^2}{3})\ln\rho+\frac{9}{4}-\pi^2-4\zeta_3\Big]\nn\\
    &&~~+C_FT_Fn_f\Big[-4\ln^2\rho+\frac{22}{9}\ln\rho+5\Big]\nn\\
    &&~~+C_FC_A\Big[11\ln^2\rho+(-\frac{169}{18}+\frac{2\pi^2}{3})\ln\rho-\frac{57}{4}+6\zeta_3\Big]\Big\}\;,\nn\\
    D_C(\rho) \label{DCform}
    &&=\frac{1}{\rho}\Big\{C_F^3\Big[-8\ln^5\rho-30\ln^4\rho+\ln^3\rho\Big(-44+8\pi^2\Big)+\ln^2\rho\Big(8\zeta_3+12\pi^2-27\Big)\\
    &&+\ln\rho\Big(-\ctworcf+48\zeta_3-\frac{41\pi^4}{90}+\frac{13\pi^2}{3}-\frac{17}{2}\Big)\nn\\
    &&+\frac{4\pi^2}{3}\zeta_3+14\zeta_3 +12\zeta_5-\frac{3\pi^4}{40}-\frac{5\pi^2}{4}-\frac{47}{8}-\frac{3}{4}\ctworcf-\frac{1}{2}\ctwozcf\Big]\nn\\
    &&+C_F^2n_fT_F\Big[\frac{40\ln^4\rho}{3}+\frac{56\ln^3\rho}{9}+\ln^2\rho\Big(-43-\frac{16\pi^2}{3}\Big)
    + \ln \rho\Big(\frac{232\zeta_3}{9}+\frac{58\pi^2}{9}-\frac{1495}{81} -\ctwornf\Big)\nn\\
    &&+\frac{254\zeta_3}{9}-\frac{7\pi^4}{15}+\frac{71\pi^2}{18}+\frac{1511}{108}
    +\frac{2}{3}\ctworcf - \frac{3}{4}\ctwornf-\frac{1}{2}\ctwoznf\Big]\nn\\
    &&+C_Fn_f^2T_F^2\Big[-\frac{112\ln^3\rho}{27}
    +\frac{68\ln^2\rho}{9}+\ln\rho\Big(\frac{140}{81}+\frac{16\pi^2}{27}\Big)
    -\frac{176\zeta_3}{27}-\frac{64\pi^2}{81}-\frac{3598}{243}+\frac{2}{3}\ctwornf\Big]\nn\\
    &&+C_FC_A^2\Big[-\frac{847\ln^3\rho}{27}+\ln^2\rho\Big(\frac{3197}{36}
    -\frac{11\pi^2}{3}\Big)+\ln\rho\Big(22\zeta_3-\frac{11\pi^4}{45}+\frac{85\pi^2}{9}-\frac{11323}{324}\Big)\nn\\
    &&-10\zeta_5
    +\frac{361\zeta_3}{27}+\frac{541\pi^4}{540}-\frac{739\pi^2}{81}
    -\frac{77099}{486}-\frac{11}{6}\ctworca\Big]\nn\\
    &&+C_F^2C_A\Big[-\frac{110\ln^4\rho}{3}+\ln^3\rho\Big(-\frac{58}{9}-\frac{8\pi^2}{3}\Big)\nn\\
    &&+\ln^2\rho\Big(-36\zeta_3+\frac{35\pi^2}{3}+\frac{467}{4}\Big)+\ln\rho\Big(-\frac{1682\zeta_3}{9}+\frac{133\pi^4}{90}
    -\frac{403\pi^2}{18}+\frac{29663}{324}
    -\ctworca\Big)\nn\\
    &&-30\zeta_5-\frac{1943\zeta_3}{18}+\frac{2\pi^2\zeta_3}{3}
    +\frac{77\pi^4}{40}-\frac{757\pi^2}{72}-\frac{49}{27}
    -\frac{11}{6}\ctworcf-\frac{3}{4}\ctworca-\frac{1}{2}\ctwozca\Big]\nn\\
    &&+C_AC_Fn_fT_F\Big[\frac{616}{27}\ln^3\rho
    +\ln^2\rho\Big(\frac{4\pi^2}{3}-\frac{512}{9}\Big)+\ln\rho\Big(8\zeta_3-\frac{128\pi^2}{27}+\frac{673}{81}\Big)\nn\\
    &&+\frac{608\zeta_3}{27}-\frac{10\pi^4}{27}+\frac{430\pi^2}{81}
    +\frac{24844}{243}-\frac{11}{6}\ctwornf+\frac{2}{3}\ctworca\Big]\}\;.
\end{eqnarray}

\section{$G_{ij}$ expansion}
Occasionally it is helpful to write an event shape distribution as
\begin{equation} \label{RserGIJ}
  R(x) = \left(1 + \sum_{m=1}^{\infty} C_m \left(\frac{\alpha}{2\pi}\right)^m\right) 
\exp \left(\sum_{i=1}^\infty \sum_{j=1}^{i+1} G_{i\,j} \left(\frac{\alpha}{2\pi}\right)^i \ln^j \frac{1}{x}\right)
 + \sum_{n=0}^\infty \alpha^n f_n(x) \,.
\end{equation}
The $G_{ij}$ and $C_m$ are calculable in SCET for exponentiation up to N${}^3$LL accuracy.

The results are
\begin{align}
    C_1&=C_F\Big(-\frac{5}{2}+\frac{\pi^2}{3}\Big)\;,\nn\\
    C_2&=C_F^2\Big(\frac{41}{8}+\frac{\pi^4}{40}-\frac{\pi^2}{2}-12\zeta_3+\frac{1}{4}\ctworcf \Big)
        +C_Fn_fT_F\Big(\frac{905}{162}-\frac{58}{9}\zeta_3-\frac{7\pi^2}{6}+\frac{1}{4}\ctwornf\Big)\nn\\
       &+C_AC_F\Big(-\frac{8977}{648} -\frac{73\pi^4}{360}+\frac{85\pi^2}{24}+\frac{481}{18}\zeta_3+\frac{1}{4}\ctworca\Big)\,,
\end{align}
and
\begin{align}
    G_{12}&=-2C_F\;,\nn\\
    G_{11}&=3C_F\;,\nn\\
    G_{23}&=C_F\Big[n_fT_F\frac{4}{3}-C_A\frac{11}{3}\Big]\;,\\
    G_{22}&=C_F\Big[-C_F\frac{2\pi^2}{3}+n_fT_F\frac{11}{9}+C_A\Big(-\frac{169}{36}+\frac{\pi^2}{3}\Big)\Big]\;,\nn\\
    G_{21}&=C_F\Big[C_F\Big(4\zeta_3+\frac{3}{4}\Big)-5n_fT_F+C_A\Big(\frac{57}{4}-6\zeta_3\Big)\Big]\;,\nn\\
    G_{34}&=C_F\Big[-C_A^2\frac{847}{108}+C_An_fT_F\frac{154}{27}-n_f^2T_F^2\frac{28}{27}\Big]\;,\nn\\
    G_{33}&=C_F\Big[C_A^2\Big(-\frac{3197}{108}+\frac{11\pi^2}{9}\Big)+n_fT_FC_A\Big(\frac{512}{27}-\frac{4\pi^2}{9}\Big)-n_f^2T_F^2\frac{68}{27}+\nn\\
    &~C_Fn_fT_F\Big(2+\frac{4\pi^2}{3}\Big)-C_FC_A\frac{11\pi^2}{3}+C_F^2\frac{16}{3}\zeta_3\Big]\;,\nn\\
    G_{32}&=C_F\Big[C_A^2\Big(11\zeta_3-\frac{11\pi^4}{90}+\frac{85\pi^2}{18}-\frac{11323}{648}\Big)+C_An_fT_F\Big(4\zeta_3-\frac{64\pi^2}{27}+\frac{673}{162}\Big)\nn\\
    &~+n_f^2T_F^2\Big(\frac{70}{81}+\frac{8\pi^2}{27}\Big)+C_F^2\Big(\frac{2\pi^4}{45}-12\zeta_3\Big)+C_FC_A\Big(-44\zeta_3+\frac{2\pi^4}{9}-\frac{239\pi^2}{108}+\frac{11}{8}\Big)\nn\\
    &~+C_Fn_fT_F\Big(8\zeta_3+\frac{13\pi^2}{27}+\frac{43}{6}\Big)\Big]\;,\nn\\
    G_{31}&=C_F\Big[C_F^2\Big(\frac{29}{8}+\pi^2-\frac{8}{3}\pi^2\zeta_3+26\zeta_3-12\zeta_5+\frac{1}{2}\ctwozcf\Big)\nn\\
    &~+C_Fn_fT_F\Big(-\frac{77}{4}+\frac{7\pi^4}{15}+\frac{11\pi^2}{9}-\frac{188}{9}\zeta_3-\frac{2}{3}\ctworcf+\frac{1}{2}\ctwoznf\Big)\nn\\
    &~+C_FC_A\Big(\frac{23}{2}-\frac{79\pi^4}{60}-\frac{175\pi^2}{36}+\frac{4\pi^2}{3}\zeta_3+\frac{493}{9}\zeta_3+30\zeta_5+\frac{11}{6} \ctworcf+\frac{1}{2}\ctwozcf\Big)\nn\\
    &~+C_A^2\Big(\frac{77099}{486}-\frac{541\pi^4}{540}+\frac{739\pi^2}{81}-\frac{361}{27}\zeta_3+10\zeta_5+\frac{11}{6}\ctworca \Big)\nn\\
    &~+C_An_fT_F\Big(-\frac{24844}{243}+\frac{10\pi^4}{27}-\frac{430\pi^2}{81}-\frac{608}{27}\zeta_3-\frac{2}{3}\ctworca+\frac{11}{6}\ctwornf \Big)\nn\\
    &~+n_f^2T_F^2\Big(\frac{3598}{243}+\frac{64\pi^2}{81}+\frac{176}{27}\zeta_3-\frac{2}{3}\ctwornf \Big)\Big]\;.
\end{align}

\end{appendix}

\end{document}